\documentclass[12pt, a4paper,margin=1.5cm]{article}
\usepackage[english]{babel}
\usepackage[utf8]{inputenc}
\usepackage{subcaption}
\usepackage{johd}
\usepackage{bm}

\usepackage{booktabs}
\usepackage{float}
\usepackage{amsmath}
\usepackage{amssymb, amsfonts}
\usepackage{adjustbox}
\usepackage[]{mathtools}
\usepackage[numbers]{natbib}
\usepackage{multirow}
    
\title{Black Hole Solutions in Non-Minimally Coupled Weyl Connection Gravity }

\author{M. Margarida Lima $^{1,2,3}$ and Cláudio Gomes $^{3,4,*}$\\
\small $^{1}$ \quad Departamento de Física, Instituto Superior Técnico, Universidade de Lisboa, \\
\small Av. Rovisco Pais,  1049-001 Lisboa, Portugal; margarida.a.lima@tecnico.ulisboa.pt\\
\small $^{2}$ \quad Centro de Análise Matemática, Geometria e Sistemas Dinâmicos, Instituto Superior Técnico, \\
\small Universidade de Lisboa, Av. Rovisco Pais, 1049-001 Lisboa, Portugal \\
\small $^{3}$ \quad OKEANOS--Instituto de Investigação em Ciências do Mar, Universidade dos Açores, \\
\small Rua Prof. Doutor Frederico Machado, 4, 9901-862 Horta, Portugal\\
\small $4$ \quad Centro de Física das Universidades do Minho e do Porto, \\
\small Rua do Campo Alegre s/n, 4169-007 Porto, Portugal\\
\small $^{*}$Corresponding author: Cláudio Gomes; \tt{claudio.fv.gomes@uac.pt} \\
}
\date{}

\begin{document}
\maketitle
\begin{abstract} 
Schwarzschild and Reissner--Nordstrøm black hole solutions are found in the context of a non-minimal matter--curvature coupling with Weyl connection both in vacuum and in the presence of a cosmological constant-like matter content. This model has the advantage of an extra force term which can mimic dark matter and dark energy, and simultaneously following Weyl's idea of unifying gravity and electromagnetism. In fact, vacuum Schwarzschild solutions differ from the ones in a constant curvature scenario in $f(R)$ theories, with the appearance of a coefficient in the term that is linear in $r$ and a corrected ``cosmological constant''. Non-vacuum Schwarzschild solutions formally have the same solutions as in the previous case, with the exception being the physical interpretation of a cosmological constant as the source of the matter Lagrangian and not a simple reparameterization of the $f(R)$ description. Reissner--Nordstrøm solutions cannot be found in a vacuum, only in the presence of matter fields, with the result that the solutions also differ from the constant curvature scenario in $f(R)$ theories by the term being linear in $r$, the corrected/dressed charge, and the cosmological constant. These results have bearings on future numerical simulations for black holes and gravitational waves in next-generation wavelet templates.

 \end{abstract}

\noindent\\

\section{Introduction}
General relativity is one of the mathematically simplest   {theories of gravity} that obey several physical and observational requirements. However, there are some problems on
both small and large scales.  At small scales, it is not compatible with quantum mechanics; hence, we still do not have a complete theory of gravity in the so-called UV regime. At~the large scale, both dark matter and dark energy are required in~order to account for the data; however, we still do not exactly know what they are made of. Additionally, there are the well-known problems of the cosmological constant and~the existence of singularities. Therefore, several alternative models to Einstein's theory have been proposed to account for astrophysical and cosmological data (see, e.g.,~\cite{Capozziello:2011,Nojiri:2011,Nojiri:2017}).

One of the most famous extensions of general relativity consists of the so-called $f(R)$ theories~\cite{Sotiriou:2010,DeFelice:2010}. A~further extension includes a non-minimal coupling between matter and curvature~\cite{Bertolami:2007}, which leads to a non-conservation law for the energy--momentum tensor built from matter fields. This feature allows for mimicking dark matter and dark energy~\cite{Bertolami:2010dm1,Bertolami:2012dm2,Bertolami:2010de}. Furthermore, this model has been explored in several astrophysical and cosmological contexts~\cite{Bertolami:2014virial,Olmo:2015,Gomes:2017inflation,Bertolami:2018gws,Ferreira:2019proc,Ferreira:2019gav,Bertolami:2020boltzmann,Gomes:2020jeans1,Gomes:2022jeans2,March:2021,Bertolami:2022magnetic,Bertolami:2023swampland}.

  {Usually, these approaches are based on a symmetric connection that is metric compatible, $\nabla_{\mu}g^{\mu\nu}=0$.} Nevertheless, there are other promising avenues, such as when looking into torsion $T$ and non-metricity $Q$. In fact, it is possible to formulate three gravity models that are equivalent to each other in general relativity where the gravity field is either   {the metric, the torsion, or related to the non-metricity.} This is not the case in theories that deviate from Einstein's theory \cite{BeltranJimenez:2019}. Likewise, for the $f(R)$ extension of general relativity, we can have $f(T)$ \cite{Linder:2010fT} and $f(Q)$ \cite{Jimenez:2018,Heisenberg:2024}.

As expected, the~non-minimal matter--curvature coupling model has been extended to its non-metricity version~\cite{Harko:2018nmcQ}, where the scalar curvature is replaced by the scalar non-metricity. Another realization involves considering that the geometry is not described by the metric field alone but also by a vector field which is related to the metric field via the non-metricity property, namely, $D_{\mu}g^{\mu\nu}=A_{\mu}g^{\mu\nu}$. This is called the Weyl connection gravity, and~is not necessarily the same as the so-called Weyl gravity, where there is a squared Weyl tensor in the action functional. Recall that this was Weyl's attempt to incorporate electromagnetism into general relativity via non-metricity, similar to the Kaluza--Klein model with an extra fifth dimension~\cite{Kaluza:1921,klein:1926}, despite criticisms from Einstein given the possibility of continuous and arbitrary length variation of a vector   {from one point to another in space--time} (except,~e.g.,~for a charge particle, the Weyl vector is purely imaginary, and, hence, problematic). Moreover, as~far as matter fields are concerned, it has been shown that local scale invariance leads to the existence of a Weyl vector meson that absorbs the Higgs particle remaining in the Weinberg--Salam model~\cite{Cheng:1988}. This Weyl vector meson, coined the metron, may interact with spinor fields under certain conditions~\cite{Aringazin:1991}. Moreover, the~sources of a general non-metricity have been shown to be the shear and dilation currents, which also couple to the former~\cite{Neeman:1997}.

The Weyl connection realization has been generalized into a non-minimal version~\cite{Gomes:2019weyl}. This model has proven to also admit cosmological solutions when the Weyl vector is dynamical and identified as a gauge vector~\cite{Baptista:2020}. In the latter scenario, it can be safe from Ostrogradsky instabilities (which arise in non-degenerate Lagrangian densities of theories that have higher order derivatives with respect to time, leading to unbounded states of energy) if either the extrinsic curvature scalar of the hypersurface of the space--time foliation is zero or if the Weyl vector has only spatial components~\cite{Baptista:2021}.

A further analysis of gravity models concerns black hole solutions and their stability. In~fact, analytic solutions in theories beyond general relativity are not trivial, and~in many cases some assumptions or simplifications are required. Thus, black hole solutions have been found by assuming constant curvature, by using perturbative methods in $f(R)$ theories~\cite{CruzDombriz:2009bhsfR,Capozziello:2010bhs,Moon:2011bhs,Rostami:2020bhs,Nashed:2019bhs}, in~the non-minimal matter--curvature coupling gravity model imposing the Newtonian limit (as in general relativity~\cite{Bertolami:2015bhs} and Weyl gravity built from the Weyl tensor \cite{Harko:2022}), or~even by looking into quasinormal modes in the latter model~\cite{Momennia:2020}. Moreover, it is also possible to study the thermodynamics of black hole solutions~\cite{Wald:2001} in modified gravity~\cite{Heisenberg:2017bhsvectortensor,Fan:2018bhsvectortensorthermodynamics,Moffat:2016bhsthermodynamics,Ghaderi:2016bhsthermodynamics,Gomes:2020bhsthermodynamics,Sebastiani:2011bhsthermodynamics}.

Thus, in~this manuscript, we aim to find black hole solutions in the context of non-minimally coupled Weyl connection~gravity.

The rest of this paper is organized as follows. In~Section~\ref{sec:model}, we present the model and some of its properties. In~the next section, we find the Schwarzschild black hole solutions both in a vacuum and assuming matter contribution in the form of a cosmological constant. In~Section~\ref{Sec:RN}, we study Reissner--Nordstrøm-like solutions in these theories. Finally, we draw our conclusions in Section~\ref{Sec:conclusions}.

\section{Non-Minimal Matter--Curvature Coupling with \\ Weyl~Connection}
\label{sec:model}

The Weyl connection gravity model introduces a vector field which provides the non-metricity properties. This model is characterized by the action of the covariant derivative of the metric field tensor not vanishing and being given by
\begin{equation}
	D_\lambda g_{\mu\nu}=A_\lambda g_{\mu\nu},
\end{equation}
where $A_\lambda$ is the Weyl vector field, $g_{\mu\nu}$ is the metric tensor, and the generalized covariant derivative is 
\begin{equation}
	D_\lambda g_{\mu\nu}=\nabla_\lambda g_{\mu\nu}-\bar{\bar{\Gamma}}^\rho_{\mu\lambda}g_{\rho\nu}-\bar{\bar{\Gamma}}^\rho_{\nu\lambda}g_{\rho\mu},
 \label{relation1}
\end{equation}
where $\nabla_\lambda$ is the usual covariant derivative with Levi--Civita connection, and $\bar{\bar{\Gamma}}^\rho_{\mu\nu}=-\frac{1}{2}\delta^\rho_{\mu}A_\nu-\frac{1}{2}\delta^\rho_\nu A_\mu+\frac{1}{2}g_{\mu\nu}A^{\rho}$ is the disformation tensor which reflects the Weyl non-metricity. Note that in contrast to the Levi--Civita part, which is not tensorial, the~disformation piece of the affine connection behaves as a~tensor. 

The Riemann tensor can be generalized in order to take the Weyl connection into account, $\bar{\Gamma}^\rho_{\mu\nu}=\Gamma^\rho_{\mu\nu}+\bar{\bar{\Gamma}}^\rho_{\mu\nu}$, such that
\begin{equation}	\bar{R}^\rho_{\mu\sigma\nu}=\partial_\sigma \bar{\Gamma}^\rho_{\nu\mu}-\partial_\nu\bar{\Gamma}^\rho_{\sigma\mu}+\bar{\Gamma}^\rho_{\sigma\lambda}
	 \bar{\Gamma}^\lambda_{\nu\mu}-\bar{\Gamma}^\rho_{\nu\lambda}\bar{\Gamma}^\lambda_{\sigma\mu}. \label{Riemann-total}
\end{equation}

By contracting the first and third indices of this generalized curvature tensor, we introduce the generalized Ricci tensor, which is given by
\begin{equation}
	\bar{R}_{\mu\nu}=R_{\mu\nu}+\frac{1}{2}A_\mu A_\nu +\frac{1}{2}g_{\mu\nu} \left(\nabla_\lambda-A_\lambda\right)A^\lambda +\tilde{F}_{\mu\nu}+\frac{1}{2}\left(\nabla_\mu A_\nu+\nabla_\nu A_\mu\right)=R_{\mu\nu}+\bar{\bar{R}}_{\mu\nu},
	\label{Riccitensor}
\end{equation}
where $R_{\mu\nu}$ is the usual Ricci tensor and $\tilde{F}_{\mu\nu}=\partial_\mu A_\nu-\partial_\nu A_\mu=\nabla_\mu A_\nu -\nabla_\nu A_\mu$ is the strength tensor of the Weyl field. It is also easy to see that the trace of the generalized Ricci tensor, that is the scalar curvature with Weyl connection, is given by
\begin{equation}
	\bar{R}=R+3\nabla_\lambda A^\lambda-\frac{3}{2}A_\lambda A^\lambda=R+\bar{\bar{R}},
\end{equation}
where $R$ is the usual Ricci~curvature.

It is known that the length norm is given by $L^2=g_{\mu\nu}dx^{\mu}dx^{\nu}$. Deriving this expression and applying the relation (\ref{relation1}), it turns out that $dL=\frac{1}{2}L A_\lambda dx^\lambda$. This leads to $L=L_0e^{\frac{1}{2}\int A_{\lambda}dx^{\lambda}}$, so the length of a vector may change from one point to another in space--time. As~$dL\geq 0$, we should impose the following constraint for the Weyl vector:
\begin{equation}
    A_\lambda dx^\lambda \geq 0. \\
    \label{contraint-Weyl}
\end{equation}

These quantities can be used to generalize the non-minimal matter--curvature coupling model~\cite{Bertolami:2007}, to incorporate the Weyl connection, whose action functional takes the form~\cite{Gomes:2019weyl}:
\begin{equation}
\label{eqn:model}
	S= \int \left(\kappa f_1(\bar{R})+f_2(\bar{R})\mathcal{L} \right)\sqrt{-g}d^4x,
\end{equation} 
where $f_1(\bar{R})$ and $f_2(\bar{R})$ are generic functions of the generalized scalar curvature, $\kappa=\frac{1}{16\pi G}$ with $G$ being the Newton's constant, $\mathcal{L}$ is the matter Lagrangian density, and $g$ is the determinant of the metric field. Throughout this article, we shall consider units such that $\kappa=1$ without loss of~generality. 
 
 Varying the action with respect to the vector field, up~to boundary terms, we obtain constraint-like equations~\cite{Gomes:2019weyl}:
\begin{equation}
 	\nabla_\lambda \Theta(\bar{R})=-A_\lambda \Theta(\bar{R}), 
 	\label{eqconstraint}
 \end{equation}
 where $\Theta(\bar{R})=F_1(\bar{R})+F_2(\bar{R})\mathcal{L}$ and $F_i(\bar{R})=\frac{d f_i(\bar{R})}{d \bar{R}}$ , $i\in \{1,2\}$. 
 
In its turn, varying the action with respect to the metric, and~taking into account the previous equation, we obtain the field equations~\cite{Gomes:2019weyl}:
\begin{equation}
 	\left( R_{\mu\nu}+\bar{\bar{R}}_{(\mu\nu)} \right) \Theta(\bar{R})-\frac{1}{2}g_{\mu\nu}f_1(\bar{R})=\frac{f_2(\bar{R})}{2}T_{\mu\nu},
 	\label{fieldeqs}
 \end{equation}
where $\bar{\bar{R}}_{(\mu\nu)}=\frac{1}{2}A_\mu A_\nu+\frac{1}{2}g_{\mu\nu} \left(\nabla_\lambda-A_\lambda\right)A^{\lambda}+\nabla_{(\mu}A_{\nu)}$ and $T_{\mu\nu}$ is the energy-momentum tensor built from the matter Lagrangian, $T_{\mu\nu}=-\frac{2}{\sqrt{-g}}\frac{\delta (\sqrt{-g}\mathcal{L})}{\delta g^{\mu\nu}}$. 

The trace of the metric field equations is
\begin{equation} 
\Theta(\bar{R}) \bar{R}-2 f_1(\bar{R})=\frac{f_2(\bar{R})}{2}T,
    \label{trace}
\end{equation}
where $T=g^{\mu\nu}T_{\mu\nu}$. 

Taking the previous relations and plugging them into Equation~(\ref{fieldeqs}), one obtains the trace-free equations:
\begin{equation}
   \Theta(\bar{R}) \left[R_{\mu\nu}-\frac{1}{4}g_{\mu\nu}R \right]+\Theta(\bar{R}) \left[\bar{\bar{R}}_{(\mu\nu)}-\frac{1}{4}g_{\mu\nu}\bar{\bar{R}} \right]=\frac{f_2(\bar{R})}{2}\left[T_{\mu\nu}-\frac{1}{4}g_{\mu\nu} T\right]. 
\label{trace-free_eqs}
\end{equation}

Note that the constraint Equation~(\ref{eqconstraint}) reduces the fourth-order theory of the usual non-minimal coupling theory into a second-order version, as~we can see in Equation~(\ref{fieldeqs}). This has the advantage of avoiding ghost instabilities, as~it has been demonstrated in Ref.~\cite{Baptista:2021}, provided that some conditions are~met. \\

Taking the divergence of the field equations, it is possible to obtain the covariant non-conservation law of the energy-momentum tensor~\cite{Gomes:2019weyl}:

\begin{equation}
	\nabla_\mu T^{\mu\nu}=\frac{2}{f_2(\bar{R})}\left[ \frac{F_2(\bar{R})}{2}\left( g^{\mu\nu}\mathcal{L}-T^{\mu\nu} \right)\nabla_\mu R+\nabla_\mu(\Theta(\bar{R}) B^{\mu\nu})-\frac{1}{2}\left(F_1(\bar{R})g^{\mu\nu}+F_2(\bar{R})T^{\mu\nu}\right)\nabla_\mu \bar{\bar{R}} \right],
	\label{nonconserveq}
\end{equation}
where $B^{\mu\nu}=\frac{3}{2}A^\mu A^\nu+\frac{3}{2}g^{\mu\nu}(\nabla_\lambda-A_\lambda)A^{\lambda}$. Thus, not only the non-minimal coupling between curvature and matter but also the non-metricity property lead to a non-trivial exchange of energy and momentum between the geometry and matter~sectors.

\subsection{Geodesic~Motion}

In order to assess the geodesics in these theories, we consider the energy-momentum tensor for a perfect fluid,
\begin{equation}
    T_{\mu\nu}=(\rho+p)u_\mu u_\nu+p g_{\mu\nu},
\end{equation}
where $\rho$ is the energy density and $p$ is the pressure. The~four-velocity, $u_\mu$, satisfies the conditions $u_\mu u^\mu =-1$ and $u^\mu \nabla_\nu u_\mu=0$. We also introduce the projection operator $h_{\mu\nu}=g_{\mu\nu}+u_\mu u_\nu$, such that $h_{\mu\nu}u^{\mu}=0$. 

Contracting Equation~(\ref{nonconserveq}) with the projection operator $h_{\lambda\nu}$, we obtain
\begin{eqnarray}
&(\rho+p)g_{\lambda\nu}u^\mu\nabla_\mu u^\nu+(\nabla_\mu p)(\delta^\mu_\lambda+u_\lambda u^\mu)=\frac{1}{f_2(\bar{R})}\left[\left(F_2(\bar{R})(\mathcal{L}-p)\nabla_\mu R - \left(F_1(\bar{R})\right.\right.\right.\nonumber \\
&\left.\left.\left. +p F_2(\bar{R}) \right)\nabla_\mu \bar{\bar{R}}\right) +\left( \delta^\mu_\lambda+u_\lambda u^\mu \right)+2\left( \nabla_\mu\left(\Theta(\bar{R}) B^\mu_\lambda\right)-u_\lambda u_\mu \nabla_\mu\left(\Theta(\bar{R})B^{\mu\nu}\right) \right) \right].
\end{eqnarray}

Finally, contracting the previous relation with $g^{\sigma\lambda}$ leads to the equation of motion for a fluid element:
\begin{equation}
    \frac{d u^\sigma}{d s}+\Gamma^\sigma_{\mu\nu}u^\mu u^\nu=f^\sigma,
    \label{geodesiceqs}
\end{equation}
where the extra force term reads as follows:
\begin{equation}
    f^\sigma= \frac{1}{(\rho+p)}\left[\frac{F_2(\bar{R})}{f_2(\bar{R})}\left( \mathcal{L}-p\right)\nabla_\mu R-\nabla_\mu p-\frac{F_1(\bar{R})+p F_2(\bar{R})}{f_2(\bar{R})}\nabla_\mu \bar{\bar{R}}+\frac{2\nabla_\nu\left(\Theta(\bar{R})B^\nu_\mu \right)}{f_2(\bar{R})} \right]h^{\mu\sigma}.
\end{equation} 

It is straightforward to check that the extra force $f^\sigma$ is orthogonal to the four-velocity of the particle, $f^\sigma u_\sigma=0$, due to the properties of the projection operator. Moreover, the~first term inside brackets arises from the non-minimal coupling and breaks the degeneracy in the Lagrangian density choice for perfects fluids that happened in general relativity~\mbox{\cite{Brown:1993,Bertolami:2008perfectfluids}}. The~second term is the same that stems from general relativity, whilst the last two terms arise from the existence of the non-metricity. In~the non-minimal matter--curvature coupling model, the~choice of $\mathcal{L}=p$ leads to a vanishing of the extra-force term in the geodesics, while in this model, a vanishing force would also require that $2\nabla_\nu\left(\Theta(\bar{R})B^\nu_\mu \right)-\left(F_1(\bar{R})+p F_2(\bar{R})\right)\nabla_\mu \bar{\bar{R}}=0$ for that~choice.

\subsection{Maxwell~Equations}

The presence of this non-minimal coupling implies that the physical implications of gravity over matter fields can be quite different from one type to another. In~particular, charged matter fields have modified dynamics. Let us consider the electromagnetic Lagrangian density
\begin{equation}
    \mathcal{L}^{^{(EM)}}=-\frac{1}{4}F_{\mu\nu}F^{\mu\nu},
\end{equation}
where $F_{\mu\nu}=\partial_\mu \Phi_\nu-\partial_\nu \Phi_\mu$ is the Faraday tensor and $\Phi_\mu$ is the electromagnetic~four-potential. 

The energy momentum tensor of the electromagnetic field is given by
\begin{equation}
T^{^{(EM)}}_{\mu\nu}=F_{\mu\alpha}F^{\alpha}_\nu - \frac{1}{4}g_{\mu\nu}F_{\alpha\beta}^{\alpha\beta}.
\label{Tensor_Eletro}
\end{equation}

When this Lagrangian is considered in the action (\ref{eqn:model}), the~variation with respect to the four-potential leads to the inhomogeneous modified Maxwell equations
\begin{equation}
    \nabla_\mu (f_2(\bar{R})F^{\mu\nu})=0.
    \label{Maxwell_eqs}
\end{equation}

As we shall see in Section~\ref{Sec:RN}, these modifications will be important when analyzing the black hole solutions with electric~charge.

\subsection{Static Spherically Symmetric~Ansatz}

In order to obtain the black hole solutions of the non-minimally coupled Weyl connection gravity model, we consider the static line element in spherical coordinates:
\begin{equation}
	ds^2=-e^{\alpha(r)}dt^2+e^{\beta(r)}dr^2+r^2(d\theta^2+\sin^2(\theta)d\phi^2),
	\label{metric}
\end{equation}
where $\alpha(r)$ and $\beta(r)$ are arbitrary functions of the distance, $r$.

Since non-rotating black holes are static spherically symmetric solutions of a gravity theory, the~vector Weyl field should not change with time. Furthermore, we do not expect this to break isotropy so its components only depend on the distance. Thus, the~vector takes the form $A_\mu=(A_0(r), A_1(r), A_2(r), A_3(r))$, where $A_i(r)$, with~$i\in\{0,1,2,3\}$, are arbitrary functions of the~distance.

Throughout this work, we seek to find and analyze solutions in a vacuum and in the presence of a ``cosmological constant"-like matter. Hence, in~both cases, the~field Equation~(\ref{fieldeqs}) implies that $\bar{\bar{R}}_{(\mu\nu)}=0$, when $\mu\neq \nu$. These relations can be converted to constraints to the Weyl vector, $A_\mu$. 

After some calculations, it is possible to conclude that there exist two types of vectors: $A_\mu=(A_0(r),A_1(r),0,0)$, such that $A_0'(r)+(A_1(r)-\alpha'(r))A_0(r)=0$; and $A_\mu=(0,A_1(r),A_2(r),0)$, such that $A_2'(r)+\left( A_1(r)-\frac{2}{r} \right)A_2(r)=0$, where the prime, ', denotes the derivative in order to r. However, it is possible to see that, in~the second case, the~field equations imply that $A_2(r)=0$. 
Thus, considering the static configuration of the problem, the~Weyl vector can only take one of the following forms: 
\begin{subequations}
\begin{align}
        A_\mu&=\left( 0,A(r),0,0 \right),\label{eqn:ansatz1}\\
        A_\mu&=\left( A_0(r),A_1(r),0,0 \right)\textrm{, with~} A_0(r)\neq 0\textrm{ and } A_0'(r)+(A_1(r)-\alpha'(r))A_0(r)=0.\label{eqn:ansatz2}  
    \end{align}
\end{subequations}

Taking into account the constraint (\ref{contraint-Weyl}), the~previous vectors must
satisfy, respectively, the~following conditions:
\begin{subequations}
\begin{align}
        A(r)&\geq 0, \forall r,\label{constr-Weyl-ansatz1}\\
        A_0(r)+A_1(r)&\geq 0, \forall r.\label{constr-Weyl-ansatz2}  
    \end{align}
\end{subequations}

Considering the most generic vector, the~relation (\ref{Riccitensor}), and  the metric (\ref{metric}), the~non-vanishing components of Ricci tensor correction are given~by 

\begin{subequations}\label{RicciCorrections}
\begin{align}
 	\bar{\bar{R}}_{00}&=-\frac{1}{2}e^{\alpha(r)-\beta(r)}\left[ A_1'(r)-A_1^2(r)+\left( \frac{3}{2}\alpha'(r)-\frac{1}{2}\beta'(r)+\frac{2}{r} \right)A_1(r) \right],\label{Rb00}\\
	\bar{\bar{R}}_{11}&=\frac{1}{2}\left[ 3 A_1'(r)+\left( \frac{1}{2}\alpha'(r)-\frac{3}{2}\beta'(r)+\frac{2}{r} \right)A_1(r)+e^{\beta(r)-\alpha(r)}A_0^2(r)\right],\label{Rb11}\\
	\bar{\bar{R}}_{22}&=\frac{1}{2}r^2 \left[e^{-\beta(r)}\left( A_1'(r)-A_1^2(r)+\left( \frac{1}{2}\alpha'(r)-\frac{1}{2}\beta'(r)+\frac{4}{r} \right)A_1(r)\right)+e^{-\alpha(r)}A_0^2(r) \right],\label{Rb22}\\
	\bar{\bar{R}}_{33}&=\sin^2(\theta)\bar{\bar{R}}_{22}.\label{Rb33}
	\end{align}
\end{subequations}

Thus, the~curvature scalar correction takes the form
\begin{equation}
	\bar{\bar{R}}=3 \left[ e^{-\beta(r)} \left( A_1'(r)-\frac{1}{2}A_1^2(r)+\left( \frac{1}{2}\alpha'(r)-\frac{1}{2}\beta'(r)+\frac{2}{r} \right)A_1(r) \right)+\frac{1}{2}e^{-\alpha(r)}A_0^2(r)\right].\label{Rb}
\end{equation}

As far as the generalized Riemann curvature tensor is concerned, its components are computed in the Appendix \ref{sec:annex}. These shall be relevant when computing the Kretschmann invariant in the next~sections.

We now proceed to obtain the black hole-like solutions of this model in the form of generalized Schwarzschild and Reissner--Nordstrøm~types.

\vspace{5cm}
\section{Schwarzschild-Like Black~Hole}
\label{Sec:Schwarzschild}
\unskip

\subsection{Vacuum}

In this section, we analyze vacuum solutions, considering the two aforementioned possible realizations for the Weyl~vector. 

In this case, the~field Equations~(\ref{fieldeqs}) take the form of a pure $f(R)$ gravity with the Weyl connection:
\begin{equation}(R_{\mu\nu}+\bar{\bar{R}}_{(\mu\nu)})F_1(\bar{R})-\frac{1}{2}g_{\mu\nu}f_1(\bar{R})=0.
\label{vacuumeqs}
\end{equation}

Taking the trace of these equations, we obtain
\begin{equation}
	\bar{R} F_1(\bar{R})-2f_1(\bar{R})=0.
\end{equation}

From this equation, two conclusions can be drawn. First of all, the~model needs to be $f_1(\bar{R})=\gamma \bar{R}^2$, with~$\gamma$ some integration constant. Secondly, we can also conclude that $\bar{R}=0$. This also occurs for usual $f(R)$ gravity~\cite{CruzDombriz:2009bhsfR}.

Please note that, when applying all these conclusions to constraint (\ref{eqconstraint}) and to relations (\ref{nonconserveq}), it is possible to see that all are trivially satisfied. Then, there are no longer any apparent restrictions on the Weyl vector, so we will analyze two different possible~scenarios. 

Considering the Ricci tensor corrections (\ref{RicciCorrections}), the~subsequent corrections for the scalar curvature (\ref{Rb}), and the field Equation~(\ref{fieldeqs}), we obtain the following: 
\begin{subequations}\label{vacuumfieldeqs}
\begin{align}
	 R_{\mu\nu}+\bar{\bar{R}}_{\mu\nu}&=0, \label{eqii}\\
     R+\bar{\bar{R}}&=0, \label{eqR} 
     \end{align}
\end{subequations}
with $\mu, \nu \in\{0,1,2\}$. 

  {We now need to explore the possible ansatz} for the Weyl vector field to solve the previous equations for the metric field free~functions.

\subsubsection{First Case: $A_\mu=\left(0,A(r),0,0\right)$}

In this subsection, we will consider the simplest Weyl vector field  $A_\mu=\left(0,A(r),0,0\right)$, with~$A(r)$ an arbitrary function. 
Comparing the time-time and radial-radial components of the Equation~(\ref{eqii}), we obtain the relation
\begin{equation}
	(1-r A(r)) \left( \alpha'(r)+\beta'(r) \right)+r(A^2(r)+2A'(r))=0.
 \label{eq25}
\end{equation}

We can further simplify this equation assuming that $\beta(r)=-\alpha(r)+\epsilon$, with~$\epsilon$ some constant. Thus, we can find a solution for $A(r)$:
\begin{equation}
	A(r)=\frac{2}{r+\omega},\label{Solution1A}
\end{equation}
where $\omega>0$, we call this the Weyl constant. We impose that $\omega$ is a positive constant to ensure that $A(r)$ is well defined and $A(r)>0$, and no singularities or null length of the parallel transported vector appear, as~can be seen in Equation~(\ref{constr-Weyl-ansatz1}). 

Thus, considering the field equations~(\ref{vacuumfieldeqs}), it easy to see that the solutions to the metric functions take the following form:
\begin{subequations}\label{solution-radialSchw}
\begin{align}
	e^{\alpha(r)}&=1-\frac{2M}{r}+\frac{2(\omega+3M)}{\omega}\frac{r}{\omega}+\frac{\omega+4M}{\omega}\left(\frac{r}{\omega}\right)^2,\label{Solution1B}\\
    e^{\beta(r)}&=\frac{1+6\left( \frac{M}{\omega} \right)}{1-\frac{2M}{r}+\frac{2(\omega+3M)}{\omega}\frac{r}{\omega}+\frac{\omega+4M}{\omega}\left(\frac{r}{\omega}\right)^2}, \label{Sotution1C}
 \end{align}
\end{subequations}
where $M>0$ is the black hole mass. We note that the relation between the components of the metric is $g_{11}=-\frac{1+6\left(\frac{M}{w}\right)}{g_{00}}$. This numerical factor can be absorbed into a redefinition of the radial variable, namely, $r\to \tilde{r}$, such that $d\tilde{r}^2=\left(1+6\left(\frac{M}{w}\right)\right)dr^2$.

Moreover, if~we are close to   {the} black hole, we have $g_{00}\approx 1-\frac{2M}{r}$, which is the same expression of the well-known Schwarzschild black hole in   {general relativity}. On~the other hand, if~we are far enough, the~metric component becomes $g_{00}\approx 1+\frac{2(\omega+3M)}{\omega}\frac{r}{\omega}+\frac{\omega+4M}{\omega}\left(\frac{r}{\omega}\right)^2$. 

It is easy to see that, performing $g_{00}=0$, we can obtain the event horizon, which occurs when $r_{H}=2M\frac{\omega}{4M+\omega}$.   {We}~can, alternatively, write $r_H=2\tilde{M}$, with~$\tilde{M}=M\frac{\omega}{\omega+4M}$.

Analyzing, globally, $A(r)$, we can see that the Weyl vector has a good behavior, in~the sense that A(r) reaches a maximum value when $r=0$, namely, $A(0)=\frac{2}{\omega}$, and~asymptotically vanishes, i.e.,~for $r \rightarrow \infty$, we have $A(r) \rightarrow 0$.

In this gravity model under study, a~Schwarzschild-like black hole can exist with a non-zero Ricci scalar. Although~the total curvature scalar is zero, the~Ricci scalar takes the form
\begin{equation}
    R=-\frac{12(r+\omega)\left( (4M+\omega)r-M\omega\right)}{\omega^2(6M+\omega)r^2}.
    \label{Ricci-sol-CASE1}
\end{equation}

\begin{figure}[H]
	\includegraphics[width=0.8\textwidth]{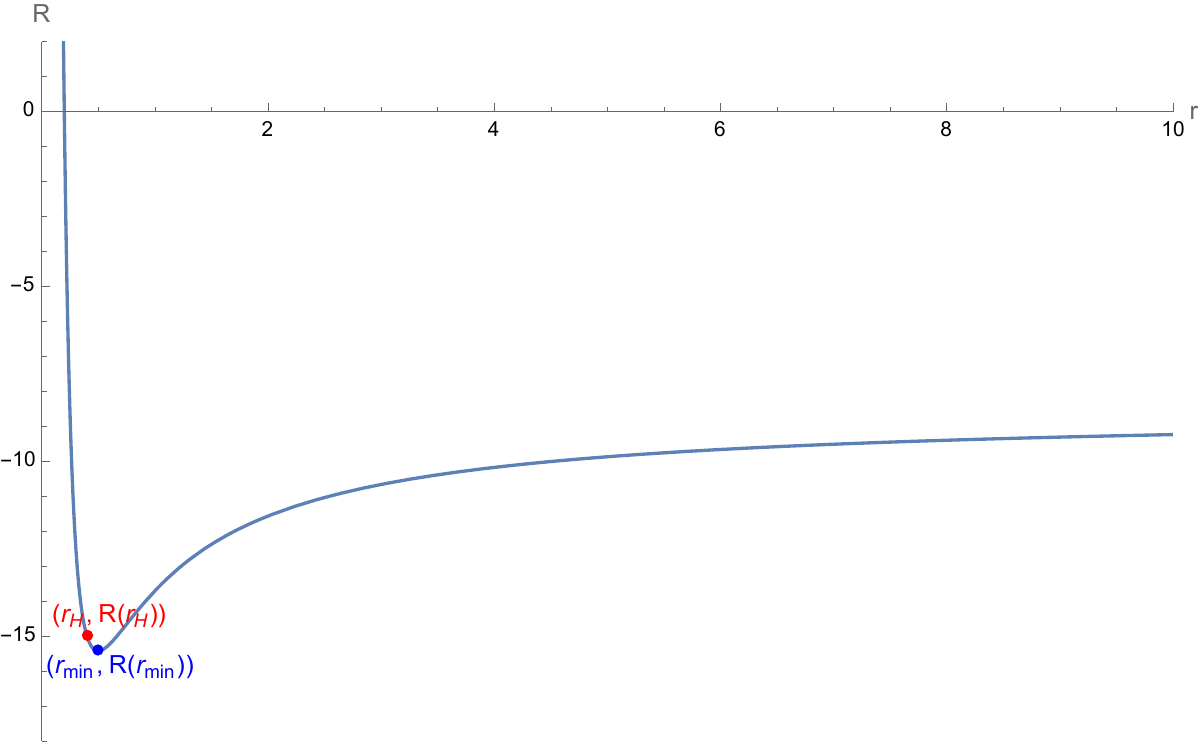} \vspace{-8pt}
	\caption{Global behavior of the standard Ricci scalar (built from the Levi--Civita part of the connection) as function of the distance, assuming $\omega=1$ and $M=1$.}\label{CASE1-Ricci}
\end{figure}

We can obtain the global behavior for the Ricci scalar graphically in Figure~\ref{CASE1-Ricci}. It is possible to see that, in~the limit $r\rightarrow\infty$, the~Ricci scalar  $R\rightarrow -\frac{12(\omega+4M)}{\omega^2(\omega+6M)}$. We can also note that the minimum value of the Ricci scalar appears for $r_\text{min}=2M\frac{\omega}{3M+\omega}$ and this occurs outside the event horizon, $r_\text{min}>r_H$. This minimum value is $R(r_\text{min})=-\frac{3(5M+\omega)^2}{M \omega^2 (6M+\omega)}$, and the Ricci scalar in the event horizon is $R(r_H)=-\frac{3(4M+\omega)}{M \omega^2}$. \\

In fact, we can assess whether the metric singularities are essential or removable by inspecting the Kretschmann invariant, which is the generalized Riemann squared $K=\bar{R}_{\mu\nu\sigma\rho}\bar{R}^{\mu\nu\sigma\rho}$, i.e.,~if those are real or due to the choice of the coordinate system. The~Kretschmann was analyzed for the Kerr--Newman black hole in general relativity in Ref.~\cite{Henry:2000}, and~for vector-tensor theories in Refs.~\cite{Campista:2020,Chan:2021}. Moreover, some gravity models are devoid of $r=0$ singularities when the atemporality mechanism is considered~\cite{Capozziello:2024}, the~Hayward regularization of Schwarzschild black holes occurs~\cite{Nashed:2024}, or~quantum metric fluctuations are present in the form of a coupling of a scalar field to the metric tensor~\cite{Yang:2021}, for~instance. In~our work, we do not pursue the latter avenues; thus, we shall compute the Kretschmann invariant to assess the nature of the singularities in our results. Thus, for~this case, we obtain
\begin{equation}
    K=\frac{48 M^2}{r^6}\left( \frac{r+\omega}{6M+\omega} \right)^2.
    \label{Kretschmann-BH1}
\end{equation}

It is possible to see that the Kretschmann invariant only diverges when $r=0$ for any $M>0$ and $\omega>0$. Therefore, $r=0$ is an essential singularity, i.e.,~a physical singularity that cannot be removed by a change of coordinates, in~the center of the black hole. At~the event horizon, the Kretschmann invariant takes the form $K_H=\frac{3(4M+\omega)^4}{4M^4\omega^4}$. Although~the curvature in the event horizon may reach a considerable absolute magnitude, the~invariant is always finite and~positive.\\

Using the relation (\ref{geodesiceqs}), we can deduce the geodesic equations. First of all, note that in this case, $f^\sigma=0$, so the geodesic equations are the usual~ones. 

Let us consider $u^{\sigma}=(t(s),r(s),0,0)$. It is possible to obtain that the trajectory described by the geodesic can be given by
\begin{equation}
    \frac{dr}{dt}=\frac{\Gamma_{00}^1 t^2+\Gamma_{11}^1 r^2}{2\Gamma_{01}^0 t r}.\label{geodesicBH1}
\end{equation}

  Considering the solution obtained in Equations~(\ref{Solution1A}) and (\ref{solution-radialSchw}),   {we can} numerically solve the previous differential equation. For~that, we will analyze three different regimes: $\omega = M$, $\omega \gg M$, and $\omega \ll M$.   {If}~$\omega \gg M$, it is possible to obtain, approximately, $g_{11}\approx-\frac{1}{g_{00}}$, $g_{00}\approx -\left( 1-\frac{2M}{r}+2\frac{r}{\omega}+\left(\frac{r}{\omega}\right)^2 \right)$. On~the other hand, if~$\omega \ll M$, it is possible to obtain the relations $g_{00}=-\left( 1-\frac{2M}{r}+6\frac{M}{\omega}\frac{r}{\omega}+4\frac{M}{\omega}(\frac{r}{\omega})^2 \right)$. Therefore, we expect that the non-metricity plays a very important~role.

\begin{figure}[H]
    \includegraphics[width=0.78\textwidth]{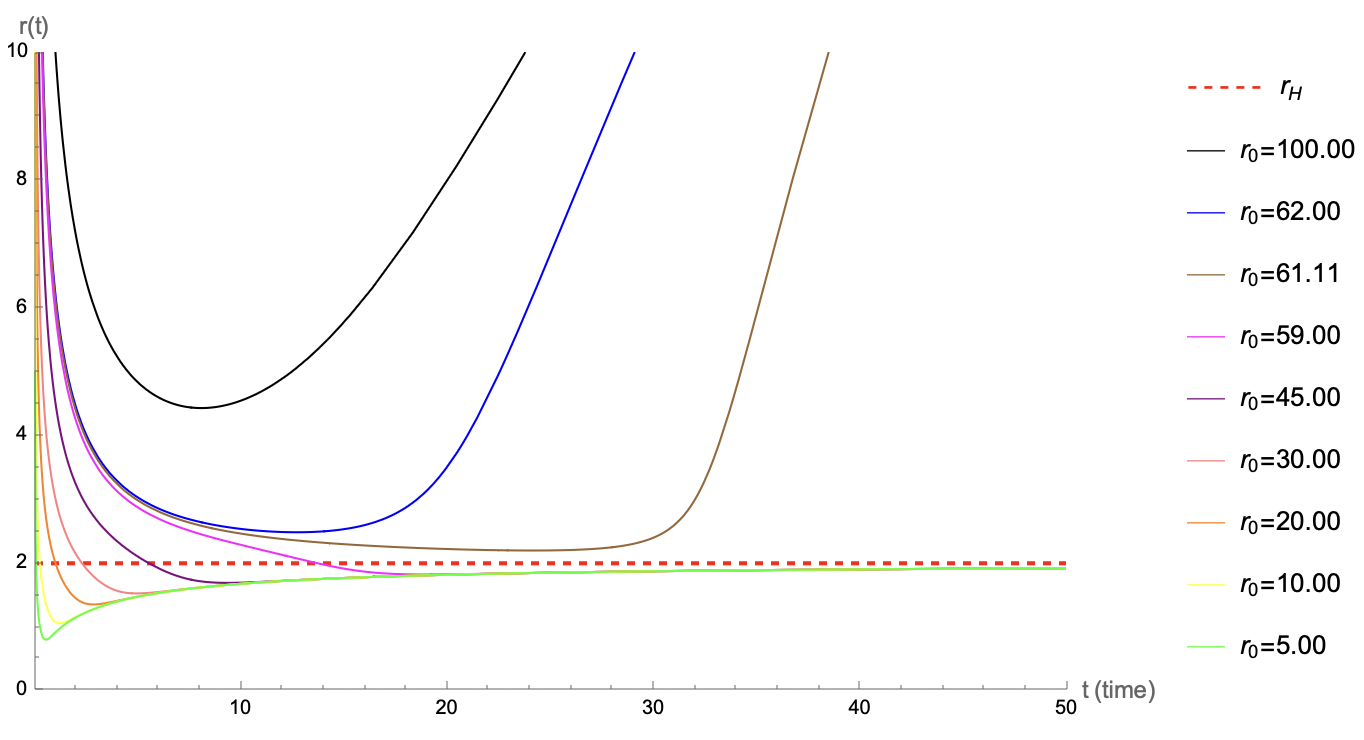}\vspace{-8pt}
    \caption{Geodesic representation of the Schwarzschild-like black hole (\ref{solution-radialSchw}), considering $\omega\gg M$, for~the parameters $\omega=10^4$ and $M=1$. For~the representation, the~initial radius is denoted by $r_0$, with~initial time $t_0=0.01$.}
    \label{fig:BH-2}
\end{figure}

\begin{figure}[H]
    \includegraphics[width=0.78\textwidth]{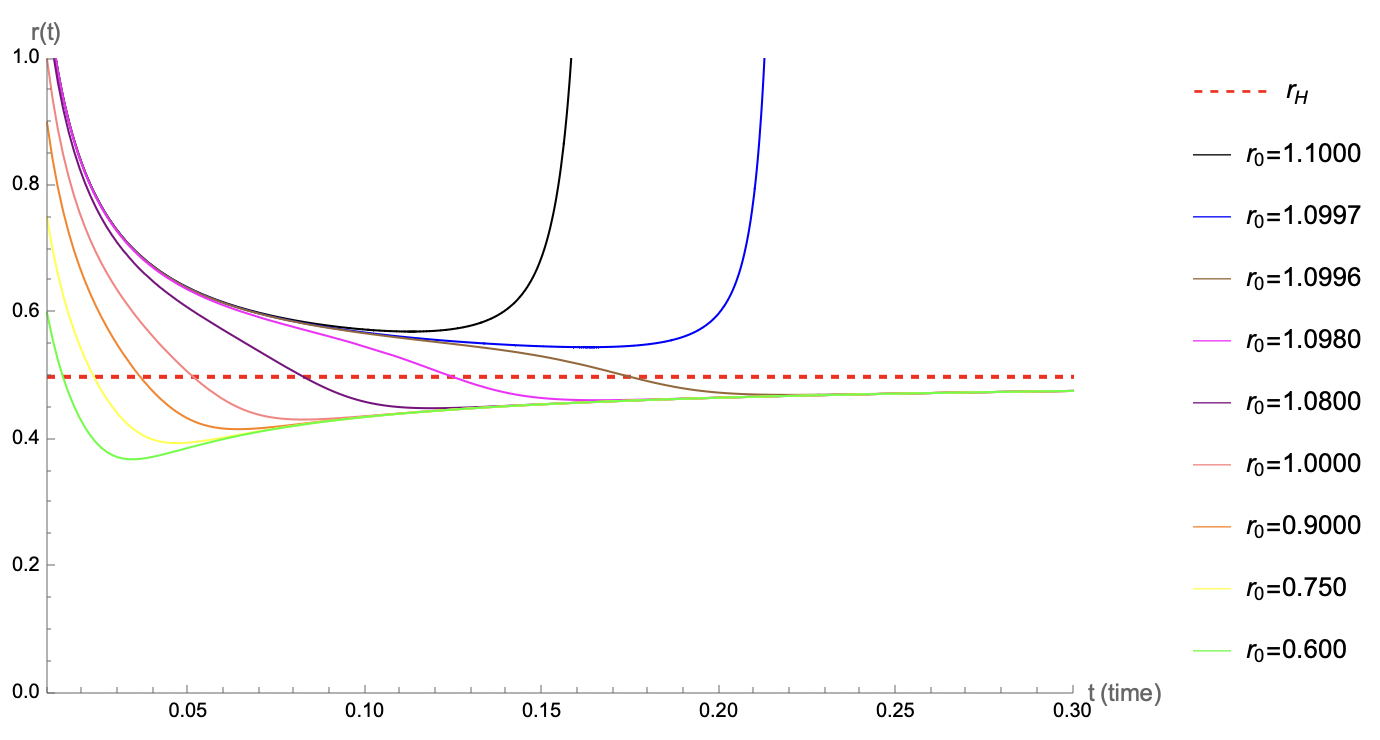}\vspace{-8pt}
    \caption{Geodesic representation of the Schwarzschild-like black hole (\ref{solution-radialSchw}), considering $\omega\ll M$, for~the parameters $\omega=1$ and $M=10^2$. For~the representation, the~initial radius is denoted by $r_0$, with~initial time $t_0=0.01$.}
    \label{fig:BH-3}
\end{figure}

These three scenarios can be explored graphically. The~numerical solutions of the geodesic equation, Equation~(\ref{geodesicBH1}), are shown in the next figures: Figure~\ref{fig:BH-1} for $\omega = M$, Figure~\ref{fig:BH-2} for $\omega \gg M$, and Figure~\ref{fig:BH-3} for $\omega \ll M$.

\begin{figure}[H]
    \includegraphics[width=0.78\textwidth]{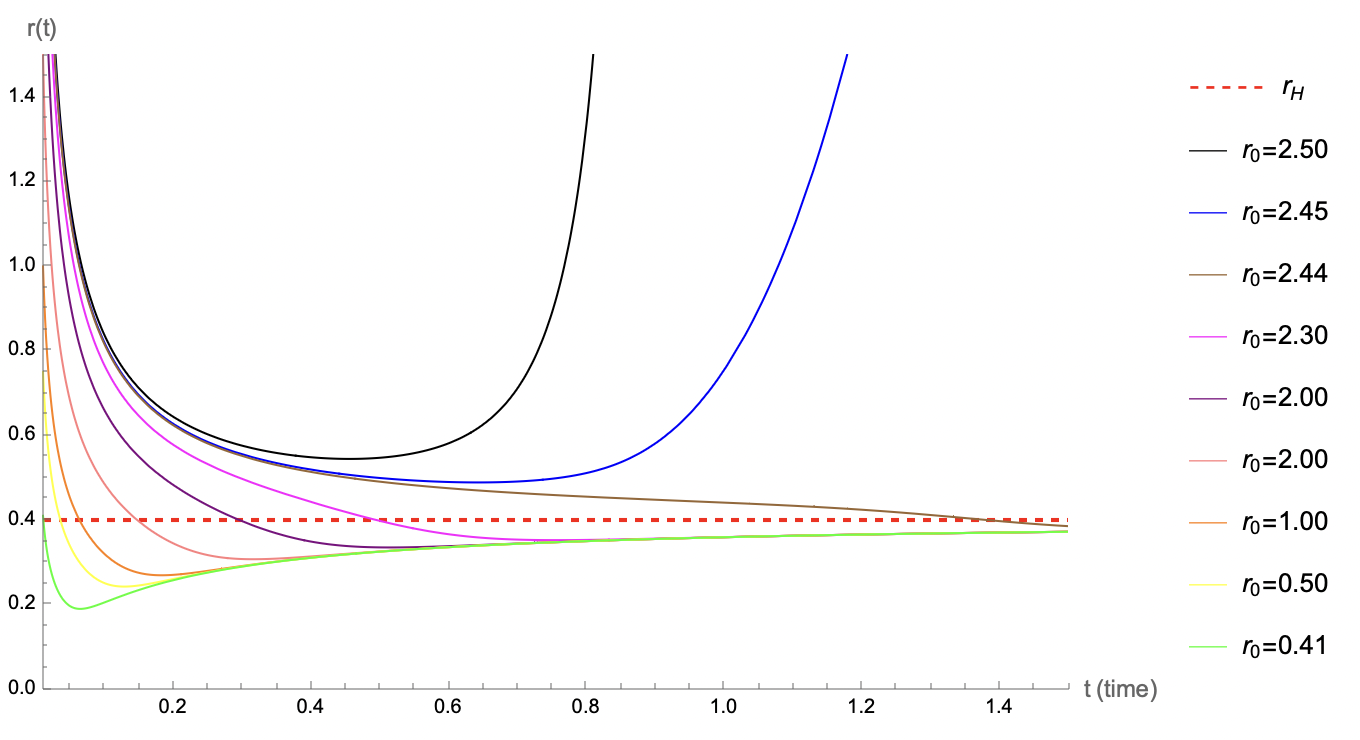}\vspace{-8pt}
        \caption{Geodesic representation of the Schwarzschild-like black hole (\ref{solution-radialSchw}), considering $\omega= M$, for~the parameters $\omega=1$ and $M=1$. For~the representation, the~initial radius is denoted by $r_0$, with~initial time $t_0=0.01$.}
        \label{fig:BH-1}
\end{figure}
\vspace{-9pt}
\vspace{-9pt}

Table~\ref{Table-BH1} represents the estimated values for the event horizon, $r_H$, and~the initial distance to the black hole (considering the initial time $t_0=0.01$), $r_{0,\text{crit}}$, after~which, for initial distance $r_0>r_{0,\text{crit}}$, the geodesic is no longer attracted to the black hole and diverges to infinity. For~that, we consider three values for the mass, $10^{-2}$, $1$, and $10^2$, and~six different values to the Weyl constant for each $M$. We now analyze the impact of each of the~parameters.

\vspace{2cm}

\begin{table}[H]
	\centering
	\footnotesize
	\begin{tabular}{|c|c|c|c|}
		\hline 
		\hspace{0.2cm} M \hspace{0.2cm} & \hspace{0.2cm} $\omega$ \hspace{0.2cm} &  $r_{H}$ & $r_{0,\text{crit}}$ \\
		\specialrule{1.5pt}{-1pt}{-1pt }
		\multirow{6}{*}{$10^{-2}$} & $10^{-3}$  & $4.87805 \times 10^{-4}$     & $4.898605 \times 10^{-4}$  \\ \cline{2-4} 
		& $10^{-2}$  & $4.0 \times 10^{-3}$     & $4.44686 \times 10^{-3}$  \\ \cline{2-4} 
		& $10^{-1}$  & $1.42857 \times 10^{-2}$     & $2.96460 \times 10^{-2}$  \\ \cline{2-4} 
		& $1$   & $1.92308 \times 10^{-2}$                    & $5.55967 \times 10^{-2}$   \\ \cline{2-4} 
		& $10^{1}$   & $1.99203 \times 10^{-2}$      & $6.07338 \times 10^{-2}$   \\ \cline{2-4} 
		& $10^{2}$   & $1.99920 \times 10^{-2}$      & $6.12471 \times 10^{-2}$   \\ 
		\specialrule{1.5pt}{-1pt}{-1pt }
		\multirow{6}{*}{$1$}  & $10^{-2}$  & $4.98753 \times 10^{-3}$                    & $2.05553 \times 10^{-3}$                 \\ \cline{2-4} 
		& $10^{-1}$  & $4.87805 \times 10^{-2}$                    & $6.97344 \times 10^{-2}$                 \\ \cline{2-4} 
		& $1$   & $0.4$                    & $2.440224$   \\ \cline{2-4} 
		& $10^{1}$   & $1.42857$      & $2.87911 \times 10^{1}$   \\ \cline{2-4} 
		& $10^{2}$   & $1.92308$      & $5.57693 \times 10^{1}$   \\ \cline{2-4} 
		& $10^{3}$   & $ 1.99203$      & $ 6.05870  \times 10^{1} $   \\  
		\specialrule{1.5pt}{-1pt}{-1pt }
		\multirow{6}{*}{$10^{2}$}  & $10^{-1}$  & $4.99875 \times 10^{-2}$      & $5.21649 \times 10^{-2}$   \\ \cline{2-4} 
		& $1$   & $4.98753 \times 10^{-1}$      & $1.09968$   \\ \cline{2-4} 
		& $10^{1}$   & $4.87805$      & $5.714214 \times 10^{1}$   \\ \cline{2-4} 
		& $10^{2}$   & $4.0 \times 10^{1}$      & $2.43480 \times 10^{3}$   \\ \cline{2-4} 
		& $10^{3}$   & $1.42857 \times 10^{2}$             & $2.87908 \times 10^{4}$              \\ \cline{2-4} 
		& $10^{4}$   & $ 1.92308 \times 10^{2} $             & $ 5.57693 \times 10^{4} $             \\ 
		\hline
	\end{tabular}
	\caption{Estimated values for the event horizon, $r_H$, and~the initial distance to the black hole, after~which the geodesic is no longer attracted to the black hole and diverges to infinity, $r_{0,\text{crit}}$, considering different values for the mass and the Weyl parameters, $M$ and $\omega$, respectively.}
	\label{Table-BH1}
\end{table}

\vspace{1cm}

\begin{figure}[H]
    \includegraphics[width=0.8\textwidth]{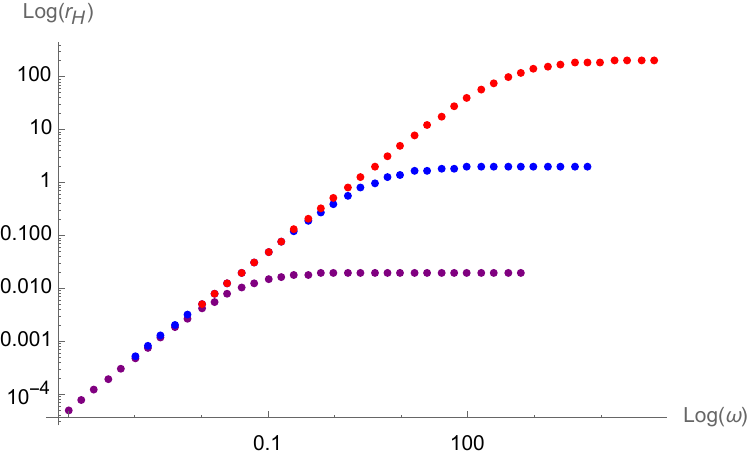}\vspace{-8pt}
    \caption{Behavior of $r_H$ compared to $\omega$, considering three different black hole masses: $M=10^{-2}$, $M=1$, and $M=10^2$, represented by the colors purple, red, and blue, respectively. To~capture all global behavior, 35 different values were considered for the $\omega$ parameter. For~$M=10^{-2}$, values were between $10^{-4}$ and $10^{3}$. For~$M=1$, values were between $10^{-3}$ and $10^{4}$. For~$M=10^{2}$, values were between $10^{-2}$ and $10^{5}$.}
    \label{fig:BH1-rH-w}
\end{figure}

\begin{figure}[H]
    \includegraphics[width=0.8\textwidth]{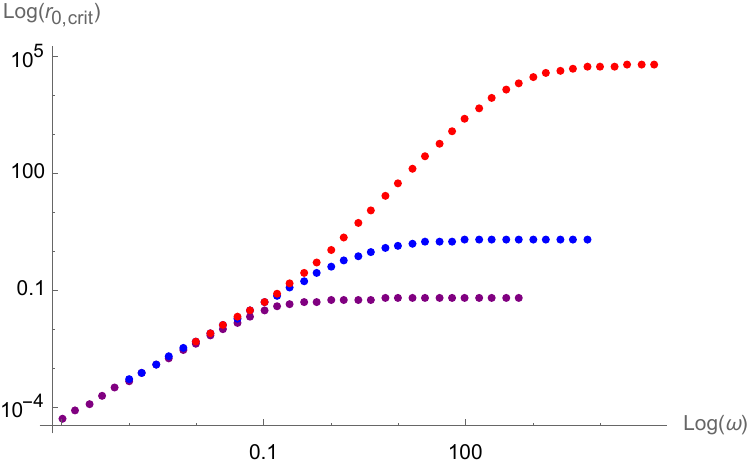}\vspace{-8pt}
    \caption{Behavior of $r_{0,\text{crit}}$ compared to $\omega$, considering three different black hole masses: $M=10^{-2}$, $M=1$, and $M=10^2$, represented by the colors purple, red, and blue, respectively. To~capture all global behavior, 35 different values were considered for the $\omega$ parameter. For~$M=10^{-2}$, values were between $10^{-4}$ and $10^{3}$. For~$M=1$, values were between $10^{-3}$ and $10^{4}$. For~$M=10^{2}$, values were between $10^{-2}$ and $10^{5}$.}
    \label{fig:BH1-r0-w}
\end{figure}

\begin{figure}[H]
    \includegraphics[width=0.8\textwidth]{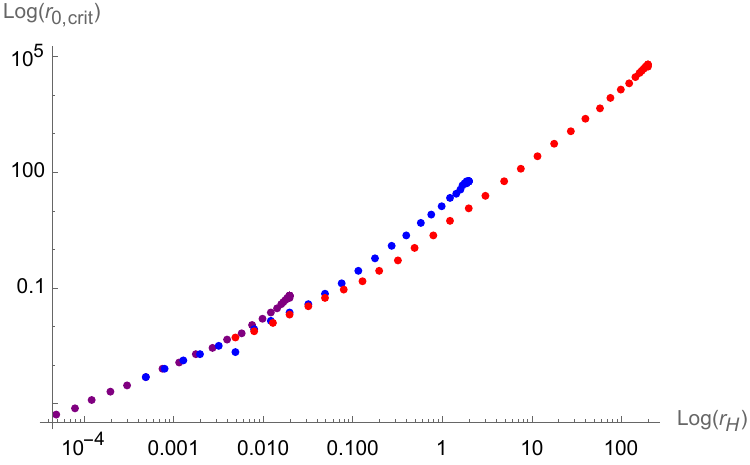}\vspace{-8pt}
    \caption{Behavior of $r_{0,\text{crit}}$ compared to $r_H$, considering three different black hole masses: $M=10^{-2}$, $M=1$, and $M=10^2$, represented by the colors purple, red, and blue, respectively. To~capture all global behavior, 35 different values were considered for the $\omega$ parameter. For~$M=10^{-2}$, values were between $10^{-4}$ and $10^{3}$. For~$M=1$, values were between $10^{-3}$ and $10^{4}$. For~$M=10^{2}$, values were between $10^{-2}$ and $10^{5}$.\\}
    \label{fig:BH1-rH-r0}
\end{figure}

In Figures~\ref{fig:BH1-rH-w}--\ref{fig:BH1-rH-r0}, we represent an analysis of the relation between the mass of the black hole, the~Weyl constant, the~results of the event horizon, and the critical radius, $r_{0,\text{crit}}$. After~that, for initial distance $r_0>r_{0,\text{crit}}$ the geodesic is no longer attracted to the black hole and diverges to infinity. For~that, 35 different values for the Weyl constant were considered, for~each of the masses, and~$r_{H}$ and $r_{0,\text{crit}}$ were calculated numerically for each of the cases. Considering this analysis, we can conclude that the global behavior of both quantities is similar regardless of the mass of the black holes (small or large) of the used values for the Weyl constant. We can also conclude that there is an asymptotic behavior in the loglog representation for the event horizon and for the critical initial value of the distance for which the geodesic is no longer attracted to the black hole, and~the higher the event horizon is, the~higher this distance is~found.
\vspace{1cm}

\vspace{-9pt}

\subsubsection{Second Case: $A_\mu=(A_0(r),A_1(r),0,0)$}\label{Subcase2}

We now consider the second possible ansatz for the Weyl vector field, namely, $A_\mu=\left(A_0(r),A_1(r),0,0\right)$, with~$A_0(r)\neq 0$ and $A_1(r)\neq 0$ being arbitrary functions that obey the constraint equation:
\begin{equation}
    A_0'(r)+(A_1(r)-\alpha'(r))A_0(r)=0.
    \label{constraintsection3.2}
\end{equation}

Using this relation to $A_1(r)$ and comparing the time--time and the radial--radial components of Equation~(\ref{eqii}), we obtain the following equation:
\begin{equation}
    2r A_0''(r)-A_0'(r)\left(r\left( \alpha'(r)+\beta'(r) \right)-4\right)=0.
\end{equation}

Analogously to the previous case, we assume that $\beta(r)=-\alpha(r)+\epsilon$, with~$\epsilon$ some constant, together with the field equations~(\ref{vacuumfieldeqs}), such that we find the following:
\begin{subequations}\label{SotulitonCase3.2}
\begin{align}
    A_0(r)&=\frac{1}{\omega}\left( 1-\frac{2M}{r}  \right),\label{Solution2A}\\
    A_1(r)&=\frac{2r}{r^2-4\omega^2}\label{Solution2B}\\
	e^{\alpha(r)}&=1-\frac{2M}{r}+\frac{M}{2\omega}\left( \frac{r}{\omega} \right)-\frac{1}{4}\left( \frac{r}{\omega} \right)^2,\label{Solution2C}\\
    e^{\beta(r)}&=\frac{1}{1-\frac{2M}{r}+\frac{M}{2\omega}\left( \frac{r}{\omega} \right)-\frac{1}{4}\left( \frac{r}{\omega} \right)^2}, \label{Sotution2D}
 \end{align}
\end{subequations}
where $M>0$ is the black hole mass and $\omega$ is the Weyl constant. We note that the relation between the time--time and radial--radial components of the metric is $g_{11}=-\frac{1}{g_{00}}$, considering, without~loss generality, $\epsilon=0$. 

It is easy to see that, by performing $g_{00}=0$, we can obtain two event horizons, which occurs when $r^{^{(M)}}_{H}=2M$ and $r^{^{(\omega)}}_{H}=2 |\omega|$. This result resembles the black hole and cosmological horizons discussion~\cite{Shankaranarayanan:2003}.

In order to understand the possible $\omega$ values, we will apply the constraint (\ref{constr-Weyl-ansatz2}) considering three different cases: $\omega=M$, $|\omega|>M$, and $|\omega|<M$. 
In the first case, considering $\omega=M$, we only have one event horizon, $R_H=2M$. The~constraint (\ref{constr-Weyl-ansatz2}) is automatically satisfied outside the event horizon for~any $M>0$. In~the second case, $|w|>M$, we have two event horizons and the external one is the $r^{^{(\omega)}}_H$. The~constraint (\ref{constr-Weyl-ansatz2}) is satisfied for any $r>r^{^{(\omega)}}_H$ if and only if $\omega>0$. Finally, when $|w|<M$, we also have two event horizons, and the external one is $r^{^{(M)}}_H$. The~constraint (\ref{constr-Weyl-ansatz2}) is satisfied for any $r>r^{^{(M)}}_H$ also if and only if $\omega>0$. Therefore, in~general, we have to impose that $\omega>0$. 

Analyzing, globally, the~Weyl vector, we can see that $A_0(r) \to \frac{1}{\omega}$ and $A_1(r)\to 0$, when $r\to \infty$. In~fact, if~we require space--time to be asymptotically flat, then $\omega \gg 1$  {;} but~if we expect that at infinity we can have non-zero background, then we see that the vector field contributes cosmologically at infinity with $A_0=1/\omega$.

In this gravity model under study, a~Schwarzschild-like black hole can exist with a non-zero Ricci scalar. Although~the total curvature scalar is zero, the~Ricci scalar takes the form
\begin{equation}
R=\frac{3(r-M)}{\omega^2 r}.
\end{equation}

\begin{figure}[H]
	\includegraphics[width=0.8\textwidth]{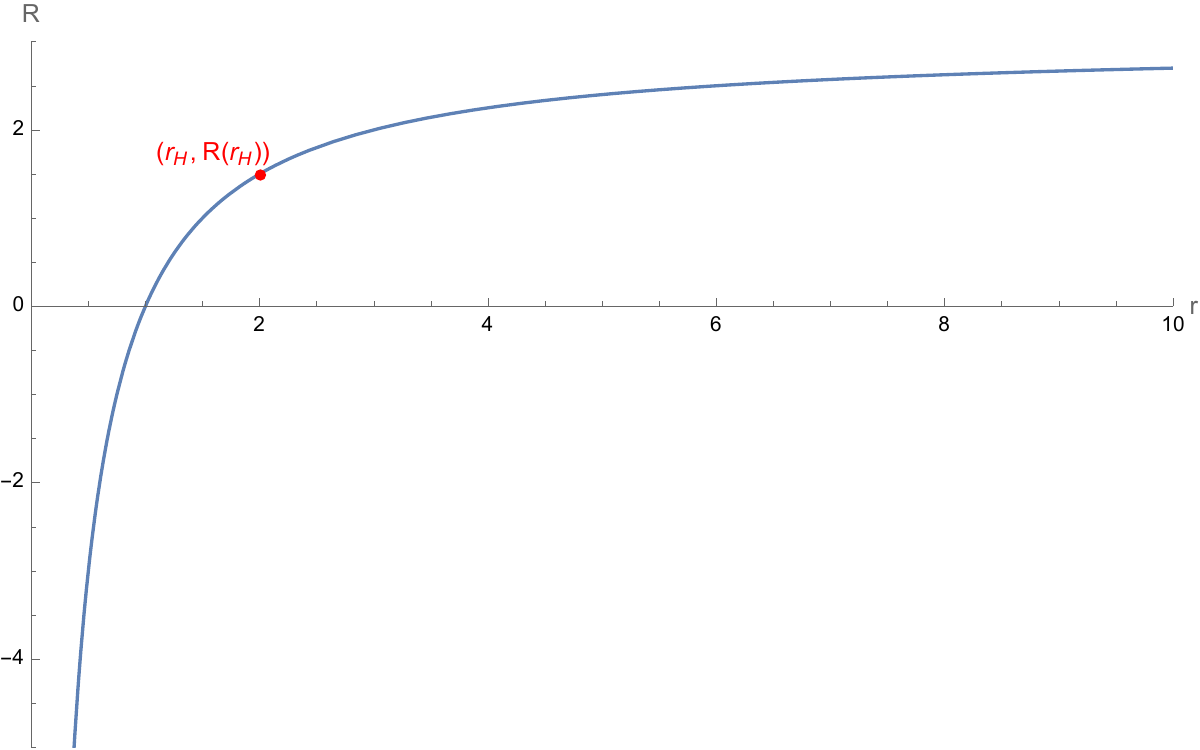} \vspace{-8pt}
	\caption{Global behavior of the standard Ricci scalar (built from the Levi--Civita part of the connection) as function of the distance, assuming $\omega=1$ and $M=1$.}\label{CASE2-Ricci}
\end{figure}

We can obtain the global behavior for the Ricci scalar graphically in Figure~\ref{CASE2-Ricci}. It is possible to see that, in~the limit $r\to \infty$, the~Ricci scalar $R\to \frac{3}{\omega^2}$. We also can note that the Ricci curvature is zero when $r=M$, i.e.,~on the surface of the black hole. When $\omega > M$, the~Ricci scalar in the external event horizon is given by $R\left(r^{^{(\omega)}}_H\right)=\frac{3(2\omega-M)}{2\omega^3}$. When $\omega \leq M$, the~Ricci scalar in the external event horizon is given by $R\left(r^{^{(M)}}_H\right)=\frac{3}{2\omega^2}$. \\

\vspace{-6pt}

In order to assess the nature of the singularities in this result, we compute the Kretschmann invariant. Thus, for~this case, we obtain
\begin{equation}
    K= \frac{48 M}{r^6}\left( \frac{4M \omega^2 (47-2M r+r^2)-r^2(44M-2r(1+M^2)+M r^2)}{192 \omega^2} \right)
    \label{Kretschmann-BH2}
\end{equation}

It is possible to see that the Kretschmann invariant only diverges when $r=0$ for any $M>0$ and $\omega>0$. Therefore, $r=0$ is an essential singularity in~the center of the black hole. At~the event horizons, the~Kretschmann invariant takes the forms $K_H^{^{(\omega)}}=\frac{M(3M+4\omega)}{64 \omega^6}$ and $K_H^{^{(M)}}=\frac{47\omega^2-40M^2}{64\omega^2 M^2}$. Although~the curvature in the event horizon may reach a considerable absolute magnitude, the~invariant is always finite. When $\omega>M$, the~invariant takes positive values at both the external, $r_H^{^{(\omega)}}$, and~the internal, $r_H^{^{(M)}}$, event horizon. When $\omega=M$, there only exists one event horizon and the invariant is also positive. When $\omega<M$, in~the internal event horizon, $r_H^{^{(\omega)}}$, the~invariant is positive. In~the external event horizon,  $r_H^{^{(M)}}$, the~invariant is positive when $\sqrt{\frac{40}{47}}M<\omega<M$, is zero when $\omega=\sqrt{\frac{40}{47}}M$, and is negative when $\omega<\sqrt{\frac{40}{47}}M$.\\
\vspace{0.5cm}

Using the relation of Equation~(\ref{geodesiceqs}), we derive the geodesics equation. Firstly, we note that in this case, $f^\sigma=0$, so the geodesics equation is the usual one from general~relativity. 

Let us further consider $u^{\sigma}=(t(s),r(s),0,0)$. Therefore, the~trajectory described by the geodesics can be given by
\begin{equation}
    \frac{dr}{dt}=\frac{\Gamma_{00}^1 t^2+\Gamma_{11}^1 r^2}{2\Gamma_{01}^0 t r}.\label{geodesicBH2}
\end{equation}

\vspace{0.5cm}

\begin{figure}[H]
    \includegraphics[width=0.78\textwidth]{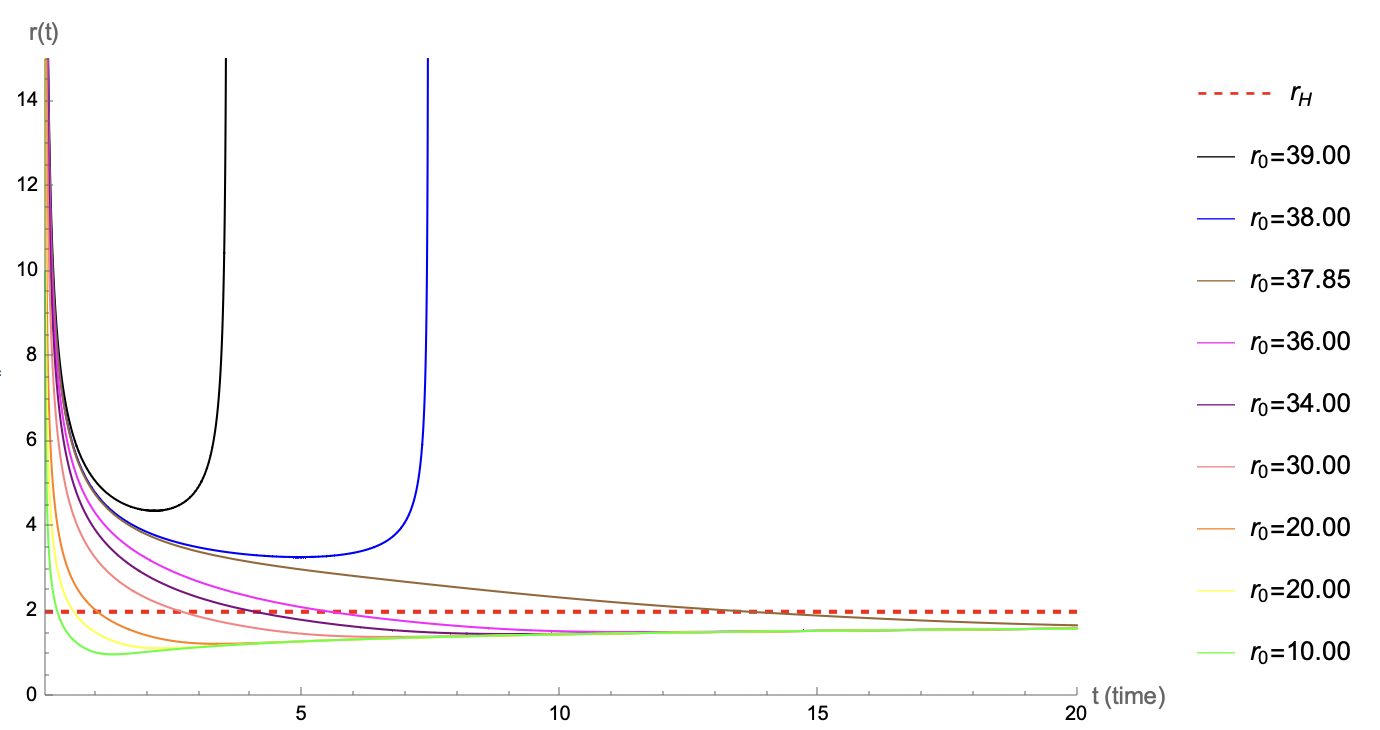}\vspace{-8pt}
    \caption{Geodesic representation of the Schwarzschild-like black hole (\ref{SotulitonCase3.2}), considering $\omega= M$, for~the parameters $\omega=1$ and $M=1$. For~the representation, the~initial radius is denoted by $r_0$, with~initial time $t_0=0.01$.}
    \label{fig:BH2-1}
\end{figure}

\begin{figure}[H]
    \includegraphics[width=0.78\textwidth]{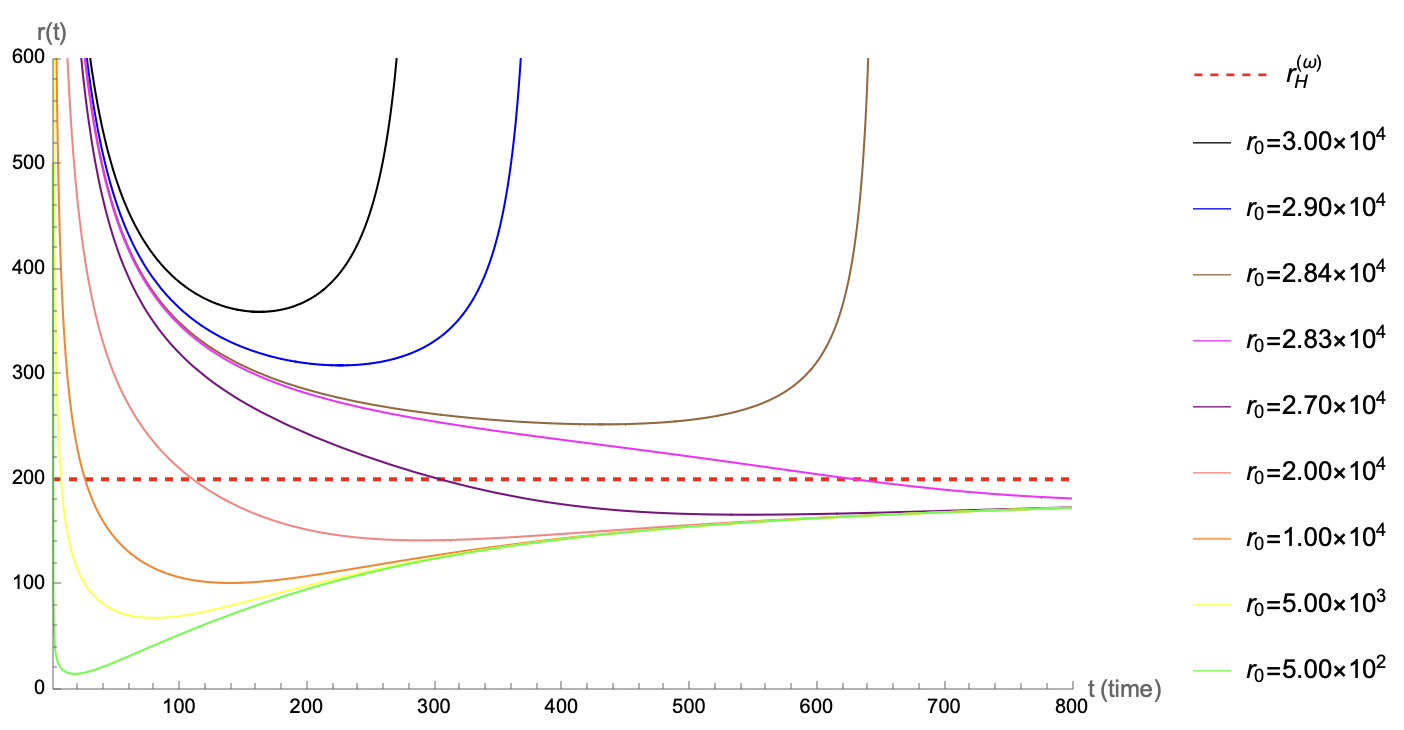}\vspace{-8pt}
    \caption{Geodesic representation of the Schwarzschild-like black hole (\ref{SotulitonCase3.2}), considering $\omega\gg M$, for~the parameters $\omega=10^2$ and $M=1$. For~the representation, the~initial radius is denoted by $r_0$, with~initial time $t_0=0.01$}
    \label{fig:BH2-2}
\end{figure}

\begin{figure}[H]
    \includegraphics[width=0.78\textwidth]{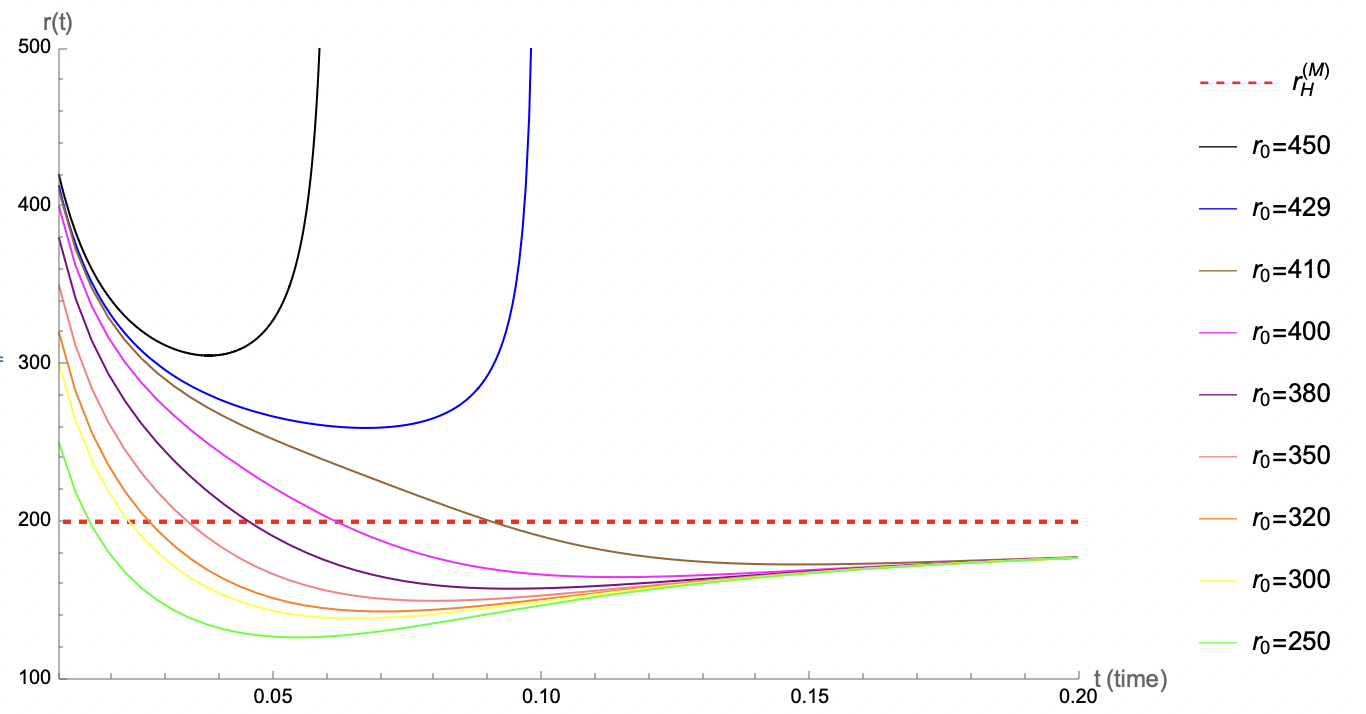}\vspace{-8pt}
    \caption{Geodesic representation of the Schwarzschild-like black hole (\ref{SotulitonCase3.2}), considering $\omega\ll M$, for~the parameters $\omega=1$ and $M=10^2$. For~the representation, the~initial radius is denoted by $r_0$, with~initial time $t_0=0.01$.}
    \label{fig:BH2-3}
\end{figure}

The~numerical solutions of the geodesic equation, Equation~(\ref{geodesicBH2}), are shown in the next figures,   considering three different scenarios: Figure~\ref{fig:BH2-1} for $\omega = M$, Figure~\ref{fig:BH2-2} for $\omega \gg M$, and Figure~\ref{fig:BH2-3} for $\omega \ll M$.

\vspace{0.5cm}

\vspace{0.5cm}
\vspace{0.5cm}

We complement the analysis with a table, namely, Table~\ref{Table-BH2}, which represents the estimated values for the external event horizon, $r_{H,\text{ext}}$ ($r_H^{^{(\omega)}}$ or $r_H^{^{(M)}}$), depending on the value of the parameters), and~the initial distance to the black hole (considering the initial time $t_0=0.01$), $r_{0,\text{crit}}$; after~that, for initial distance $r_0>r_{0,\text{crit}}$, the geodesic is no longer attracted to the black hole and diverges to infinity. For~that, we consider three types of masses, $10^{-2}$, $1$, and $10^2$, and~six different values to the Weyl constant for each $M$. Here, we intend to analyze the impact of each of the~parameters.

\begin{table}[H]
	\centering
	\footnotesize
	\begin{tabular}{|c|c|c|c|}
		\hline 
		\hspace{0.2cm} M \hspace{0.2cm} & \hspace{0.2cm} $\omega$ \hspace{0.2cm} &  $r_{H,\text{ext}}$ &  $r_{0,\text{crit}}$ \\
		\specialrule{1.5pt}{-1pt}{-1pt }
		\multirow{6}{*}{$10^{-2}$} & $10^{-3}$  & $2.0 \times 10^{-2}$     & $2.03962 \times 10^{-2}$  \\ \cline{2-4} 
		& $10^{-2}$  & $2.0 \times 10^{-2}$     & $4.74587 \times 10^{-2}$  \\ \cline{2-4} 
		& $10^{-1}$  & $2.0 \times 10^{-1}$     & $9.40337 \times 10^{-1}$  \\ \cline{2-4} 
		& $1$   & $2.0$                    & $2.84252 \times 10^{1}$   \\ \cline{2-4} 
		& $10^{1}$   & $2.0 \times 10^{1}$      & $8.94872 \times 10^{2}$   \\ \cline{2-4} 
		& $10^{2}$   & $2.0 \times 10^{2}$      & $2.82857 \times 10^{4}$   \\ 
		\specialrule{1.5pt}{-1pt}{-1pt }
		\multirow{6}{*}{$1$}  & $10^{-2}$  & $2.0$                    & $2.03925$                 \\ \cline{2-4} 
		& $10^{-1}$  & $2.0$                    & $4.12739$                 \\ \cline{2-4} 
		& $1$   & $2.0$                    & $3.79018 \times 10^{1}$   \\ \cline{2-4} 
		& $10^{1}$   & $2.0 \times 10^{1}$      & $9.17359 \times 10^{2}$   \\ \cline{2-4} 
		& $10^{2}$   & $2.0 \times 10^{2}$      & $2.83548 \times 10^{4}$   \\ \cline{2-4} 
		& $10^{3}$   & $2.0 \times 10^{3}$      & $8.94650 \times 10^{5}$   \\  
		\specialrule{1.5pt}{-1pt}{-1pt }
		\multirow{6}{*}{$10^{2}$}  & $10^{-1}$  & $2.0 \times 10^{2}$      & $2.03924 \times 10^{2}$   \\ \cline{2-4} 
		& $1$   & $2.0 \times 10^{2}$      & $4.12172 \times 10^{2}$   \\ \cline{2-4} 
		& $10^{1}$   & $2.0 \times 10^{2}$      & $3.41783 \times 10^{3}$   \\ \cline{2-4} 
		& $10^{2}$   & $2.0 \times 10^{2}$      & $3.77465 \times 10^{4}$   \\ \cline{2-4} 
		& $10^{3}$   & $2.0 \times 10^{3}$             & $9.17119 \times 10^5$              \\ \cline{2-4} 
		& $10^{4}$   & $2.0 \times 10^{4}$             & $2.83541 \times 10^{7}$              \\ 
		\hline
	\end{tabular}
	\caption{Estimated values for the external event horizon, $r_{H,\text{ext}}$, and~the initial distance to the black hole, after~which the geodesic is no longer attracted to the black hole and diverges to infinity, $r_{0,\text{crit}}$, considering different values for the mass and the Weyl parameters, $M$ and $\omega$, respectively.}
	\label{Table-BH2}
\end{table}

\vspace{0.5cm}

\begin{figure}[H]
    \includegraphics[width=0.8\textwidth, trim=0cm 0cm 0cm 0cm, clip]{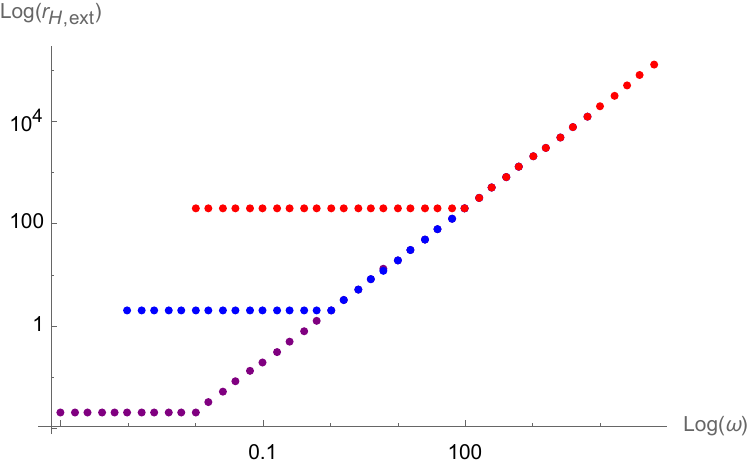}\vspace{-8pt}
    \caption{Behavior of $r_{H,\text{ext}}$ compared to $\omega$, considering three different black hole masses: $M=10^{-2}$, $M=1$, and $M=10^2$, represented by the colors purple, red, and blue, respectively. To~capture all global behavior, 35 different values were considered for the $\omega$ parameter. For~$M=10^{-2}$, values were between $10^{-4}$ and $10^{3}$. For~$M=1$, values were between $10^{-3}$ and $10^{4}$. For~$M=10^{2}$, values were between $10^{-2}$ and $10^{5}$.}
    \label{fig:BH2-rH-w}
\end{figure}

\begin{figure}[H]
    \includegraphics[width=0.8\textwidth, trim=0cm 0cm 0cm 0cm, clip]{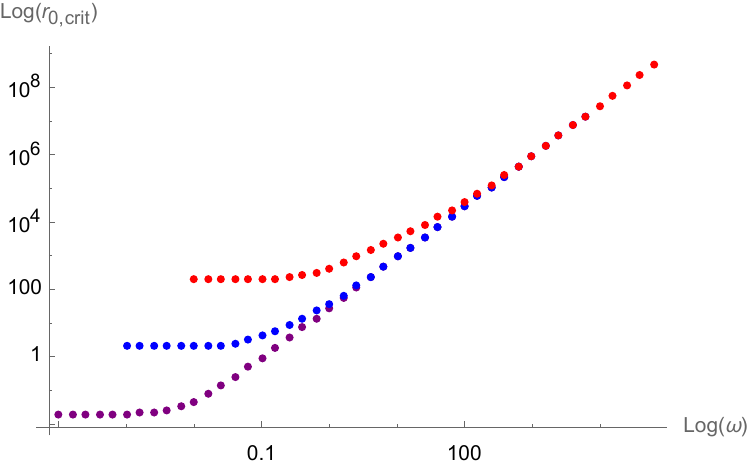}\vspace{-8pt}
    \caption{Behavior of $r_{0,\text{crit}}$ compared to $\omega$, considering three different black hole masses: $M=10^{-2}$, $M=1$, and $M=10^2$, represented by the colors purple, red, and blue, respectively. To~capture all global behavior, 35 different values were considered for the $\omega$ parameter. For~$M=10^{-2}$, values were between $10^{-4}$ and $10^{3}$. For~$M=1$, values were between $10^{-3}$ and $10^{4}$. For~$M=10^{2}$, values were between $10^{-2}$ and $10^{5}$.}
    \label{fig:BH2-r0-w}
\end{figure}

\begin{figure}[H]
    \includegraphics[width=0.8\textwidth, trim=0cm 0cm 0cm 0cm, clip]{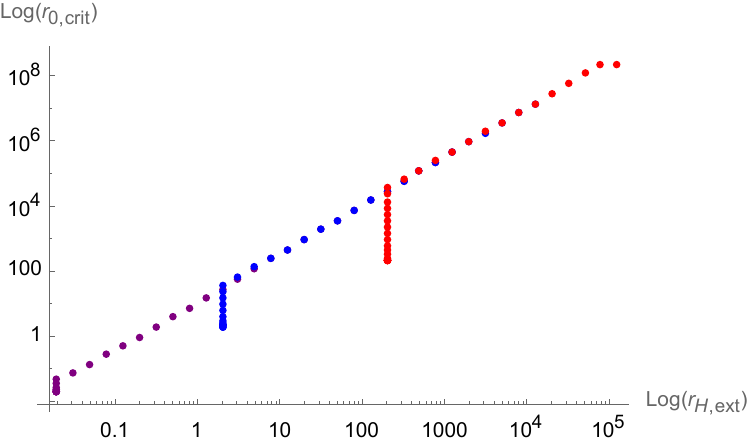}\vspace{-8pt}
    \caption{Behavior of $r_{H,\text{ext}}$ compared to $r_{0,\text{crit}}$, considering three different black hole masses: $M=10^{-2}$, $M=1$, and $M=10^2$, represented by the colors purple, red, and blue, respectively. To~capture all global behavior, 35 different values were considered for the $\omega$ parameter. For~$M=10^{-2}$, values were between $10^{-4}$ and $10^{3}$. For~$M=1$, values were between $10^{-3}$ and $10^{4}$. For~$M=10^{2}$, values were between $10^{-2}$ and $10^{5}$.\\}
    \label{fig:BH2-rH-r0}
\end{figure}

In Figures~\ref{fig:BH2-rH-w}--\ref{fig:BH2-rH-r0}, we represent an analysis of the relation between the mass of the black hole,  the~Weyl constant, the~results of the event horizon, and the critical radius, $r_{0,\text{crit}}$; after~that, for initial distance $r_0>r_{0,\text{crit}}$, the geodesic is no longer attracted to the black hole and diverges to infinity. For~that, we considered 35 different values for the Weyl constant, for~each of the masses, and~$r_{H,\text{ext}}$ and $r_{0,\text{crit}}$ were calculated numerically for each of the cases. Due to the limitations of numerical simulations, we cannot plot much further (higher values for $\omega$); therefore, we cannot directly compare   {to} the general behavior from the first ansatz for the Weyl vector field. We note, however, that this case has two horizons (possibly a black hole and a cosmological ones; see, e.g.,~Refs.~\cite{Shankaranarayanan:2003,Moffat:2016bhsthermodynamics}).

\unskip


\subsection{Cosmological Constant~Background}

In this section, we will analyze the black hole solutions when the matter Lagrangian density is non-vanishing. For~simplicity and in order to grasp the first non-trivial solution, we shall consider it in the form of a cosmological constant, i.e.,~a constant energy density. Considering $\mathcal{L}^{^{(\Lambda)}}=-2 \Lambda$, the~energy-momentum tensor components are given by
\begin{equation}
    T^{^{(\Lambda)}}_{\mu\nu}=-2 \Lambda g_{\mu\nu}. 
\end{equation}

The field Equation~(\ref{fieldeqs}) takes the form
\begin{equation}
    \bar{R}_{\mu\nu} \left( F_1(\bar{R})-2\Lambda F_2(\bar{R})\right)-\frac{1}{2} \left(f_1(\bar{R})-2\Lambda f_2(\bar{R}) \right)=0.
\end{equation}

Taking the trace of these equations, we obtain the relation
\begin{equation}
    f_1(\bar{R})-2\Lambda f_2(\bar{R})=\gamma \bar{R}^2+\xi,
    \label{constrantfunctions}
\end{equation}
where $\gamma$ and $\xi$ are~constants. 

Thus, the~field equations are
\begin{equation}
    \bar{R}_{\mu\nu}-\frac{1}{4}g_{\mu\nu}\bar{R}=0, 
\end{equation}
that implies that
\begin{eqnarray}
    &&\bar{R}_{\mu\nu}=0,\\ \label{eqRij}
    &&\bar{R}=0. \label{eqRbar}
\end{eqnarray}

It is straightforward to verify that, taking Equations~(\ref{constrantfunctions}) and~(\ref{eqRbar}) into account, both constraint Equation~(\ref{eqconstraint}) and the non-conservation law (\ref{nonconserveq}) are automatically satisfied for~any vector field $A_\lambda$. Therefore, the vacuum solutions found above are also solutions when we consider a cosmological constant background. Therefore, we can interpret this result as stemming from the mathematical reparametrization, namely, $f(\bar{R})=f_1(\bar{R})-2\Lambda f_2(\bar{R})$ in Equation~(\ref{constrantfunctions}). However, as~discussed in Ref.~\cite{Bertolami:2015bhs}, physically, these two models are distinct, as one has a physical meaning for the cosmological constant-like Lagrangian (the energy from vacuum). Thus, we can relate this ``cosmological constant'' with a cosmological constant appearing as an integration constant of the field equations. The~non-minimal coupling model has an advantage of allowing for contributing   {to} an explanation of the so-called cosmological constant~problem.\\

\subsection{Thermodynamics}
\label{Sec:thermodynamics}

We now evaluate some thermodynamics quantities for the black hole solutions found in this paper for the non-minimally coupled Weyl connection gravity model. We closely follows Refs.~\cite{Gomes:2020bhsthermodynamics,Sebastiani:2011bhsthermodynamics,Moffat:2016bhsthermodynamics} in order to compute such~quantities.

Therefore, for~the metric (\ref{metric}), the~black hole temperature from the quantum tunneling method (which is equivalent to the Hawking method~\cite{Gomes:2020bhsthermodynamics}, and~we denote by the superscript ``BH'') is given by
\begin{equation}
T^{^{(BH)}}=\frac{\sqrt{\left(e^{\alpha(r_{H,\text{ext}})}\right)'\left(e^{-\beta(r_{H,\text{ext}})}\right)'}}{4\pi}=\frac{\sqrt{-\alpha'(r_{H,\text{ext}})\beta'(r_{H,\text{ext}}) e^{\alpha(r_{H,\text{ext}})-\beta(r_{H,\text{ext}})}}}{4\pi}.
    \label{temperature-eq}
\end{equation}

The black hole entropy is given by
\begin{equation}
    S^{^{(BH)}}=\int \frac{dM}{T^{^{(BH)}}(M)}
\label{entropy-eq}
\end{equation}
and the black hole heat capacity at constant volume takes the form
\begin{equation}
    C_V^{^{(BH)}}=T^{^{(BH)}} \frac{\partial S^{^{(BH)}}}{\partial T^{^{(BH)}}}=T^{^{(BH)}} \frac{\partial S^{^{(BH)}}}{\partial r_{H,\text{ext}}} \left( \frac{\partial T^{^{(BH)}}}{\partial r_{H,\text{ext}}} \right)^{-1}.
    \label{heat-capacity-eq}
\end{equation}

We can also derive such quantities from the so-called ``area approach''. Thus, the~area temperature reads as follows:
\begin{equation}
    T^{^{(A)}}=\frac{r_{H,\text{ext}}-r_{H,\text{int}}}{4\pi r_{H,\text{ext}}^2}.
    \label{temperature-area}
\end{equation}

Analogously, the~area entropy is
\begin{equation}
    S^{^{(A)}}=\frac{A(r_{H,\text{ext}})}{4},
    \label{entropy-area}
\end{equation}
and the area specific heat capacity is
\begin{equation}
    C_V^{^{(A)}}=T^{^{(A)}} \frac{\partial S^{^{(A)}}}{\partial T^{^{(A)}}}=T^{^{(A)}} \frac{\partial S^{^{(A)}}}{\partial r_{H,\text{ext}}} \left( \frac{\partial T^{^{(A)}}}{\partial r_{H,\text{ext}}} \right)^{-1}.
    \label{heat-capacity-area}
\end{equation}

When both definitions do not coincide, then we need to modify the first law of thermodynamics~\cite{Ma:2014,Maluf:2018,Gomes:2020bhsthermodynamics}; likewise, the non-minimal coupling model has bearings on the second law~\cite{Bertolami:2020boltzmann}, such that
\begin{equation}
    dM=TdS \implies T^{^{(A)}}dS^{^{(A)}}=T^{^{(BH)}}dS^{^{(BH)}},
\end{equation}
from which one can parametrize the following:
\begin{equation}
    T^{^{(BH)}}dS^{^{(BH)}}=L^{^{(A)}} dM,
\end{equation}
where $L^{^{(A)}}=T^{^{(A)}}\frac{dS^{^{(A)}}}{dM}$ is a function to be determined~\cite{Gomes:2020bhsthermodynamics}.

Finally, for~both definitions, the~free energy stems from
\begin{equation}
    F=M-T S.
    \label{free-energy}
\end{equation}

We now apply these definitions to each solution found for the Schwarzschild-like black~holes.

\subsubsection{Schwarzschild-Like Black Hole: First~Case}

Considering the solution (\ref{solution-radialSchw}), the~previous thermodynamical quantities read as follows:
\begin{subequations}\label{quanntities-Schw-case1}
\begin{align}
T^{^{(BH)}} &=T_0 \left(1+6\frac{M}{\omega}\right)^{-3/2},\label{temperature-BH-Schw1}\\
S^{^{(BH)}} &=\frac{8\pi}{9}\left(1+6\frac{M}{\omega}\right)^{-3/2}(3M+\omega)(6M+\omega),\label{entropy-BH-Schw1}\\
C_V^{^{(BH)}}&=C_0 \left(1+6\frac{M}{\omega}\right)^{-3/2} \frac{6M+\omega}{\omega-3M},\label{heat-BH-Schw1}   
\end{align}
\end{subequations}
where $T_0=\frac{1}{8\pi M}$ and $C_0=-8\pi M^2$ are the Hawking temperature  and the heat capacity at constant volume for the usual Schwarzschild black hole, respectively. This expression for the specific heat means that the black hole is stable for $\omega>3M$, and~unstable for $\omega<3M$, as the black hole will evaporate, leaving a stable cold remnant, in~contrast with standard Schwarzschild black holes in general relativity, which are always unstable as their temperatures increase as they absorb mass. The~specific heat diverges at $\omega=3M$ despite not corresponding to the maximum temperature, which is a decreasing function of the mass and has a maximum at zero mass, i.e.,~$T_\text{max}=T_0$. Nonetheless, this might signal the existence of a thermodynamical critical transition~point.

Using the relation (\ref{free-energy}), the~free energy considering the tunneling method takes the form
\begin{equation}
    F^{^{(BH)}}=-\left( M+\omega+\frac{\omega^2}{9M} \right),
\end{equation}
which can be minimized when $\omega=0$ and $M\neq 0$. Thus, the~system in the quantum tunneling description can be globally stable, as~it allows for non-vanishing masses to minimize its free~energy.

We now proceed to compute the previous quantities using the area method. We draw our attention to the fact that there is only one event horizon, namely, 
$r_{H}=2M\frac{\omega}{4M+\omega}$, thus being formally equivalent to impose $r_{H,int}=0$.

Therefore, the~area temperature, the~area entropy, and the area specific heat capacity at constant volume are given by
\begin{subequations}
\begin{align}
        T^{^{(A)}}&=T_0\frac{4M+\omega}{\omega},\\
        S^{^{(A)}}&=4\pi \left(\frac{M\omega}{4M+\omega}\right)^2,\\
        C_V^{^{(A)}}&=C_0 \left(\frac{\omega}{4M+\omega}\right)^2  ~.\label{heat-area-Schw1}
    \end{align}
\end{subequations}

As we can see, the~area specific heat capacity at constant volume is always positive definite; thus, the system is locally stable for all values of the mass and of the Weyl constant. Furthermore, using the relation (\ref{free-energy}), the~free energy considering the ``area approach'' takes the form
\begin{equation}
    F^{^{(A)}}=\frac{M\left( 8M+\omega \right)}{2\left( 4M+\omega \right)},
\end{equation}
which is minimized for $M=0$; thus, the system is globally unstable in the area~method.

As the area and quantum tunneling methods lead to different results, the first law must be revisited~\cite{Gomes:2020bhsthermodynamics}. In~this case, the~function $L^{^{(A)}}=\frac{\omega^2}{(4M+\omega)^2}$.

\subsubsection{Schwarzschild-Like Black Hole: Second~Case}

  {Taking into account} the solution (\ref{SotulitonCase3.2}) and applying the expression (\ref{temperature-eq}), considering the external event horizon, it is possible to see that $T^{^{(BH)}}$ is purely imaginary and diverges, which is non-physical:
\begin{equation}
    T^{^{(BH)}}=\frac{i}{2\pi r} \left| \frac{r^3-M(r^2+4\omega^2)}{(r-2M)(r^2-4\omega^2)} \right|. 
\end{equation}


If we instead compute the thermodynamical quantities via the ``area approach'', we have three cases to analyze.

The first case corresponds to $\omega < M$, for~which the two event horizons read $r_{H,\text{int}}=2\omega$ and $r_{H,\text{ext}}=2M$. Thus, the~area temperature, the~area entropy, and the specific heat at constant volume are
\begin{subequations}
\begin{align}
        T^{^{(A)}}&=T_0 \left(1-\frac{\omega}{M}\right),\\
        S^{^{(A)}}&=4\pi M^2,\\
        C_V^{^{(A)}}&=C_0 \frac{\omega-M}{2\omega-M}~.
    \end{align}
\end{subequations}

As we can see, the~area temperature is always positive and does not diverge, and~the area specific heat is positive when $\omega<M<2\omega$, negative when $2\omega<M$, and diverges at $M=2\omega$. This critical point does not correspond to the maximum value of the temperature that occurs for a vanishing Weyl constant and has the value of $T_\text{max}=T_0$. Nevertheless, one can see that the system is locally unstable for negative values of the specific heat, and~stable for positive ones. Therefore, the~free energy is given by
\begin{equation}
    F^{^{(A)}}=\frac{\omega+M}{2},
\end{equation}
which is minimized for both vanishing mass and Weyl~constant.

As for the second case, namely, when $\omega > M$, we have $r_{H,\text{int}}=2M$ and $r_{H,\text{ext}}=2\omega$. Thus, the~area temperature, entropy, and specific heat are given by
\begin{subequations}
\begin{align}
        T^{^{(A)}}&=T_0 \frac{M}{\omega} \left(1-\frac{M}{\omega}\right),\\
        S^{^{(A)}}&=4\pi \omega^2,\\
        C_V^{^{(A)}}&=C_0 \left( \frac{\omega}{M} \right)^2 \frac{M-\omega}{2M-\omega}.
    \end{align}
\end{subequations}

In this case, the~temperature is always positive and does not diverge, and~the specific heat is positive when $M<\omega<2M$ and this Shwarzschild-like black hole evaporates to a locally stable cold remnant, negative when $2M<\omega$, and diverges for $\omega=2M$. In~this case, the~critical point of the specific heat corresponds to the   {temperature maximum}, $T_\text{max}=\frac{T_0}{4}$. Finally, the~free energy is given by
\begin{equation}
    F^{^{(A)}}=\frac{3M-\omega}{2},
\end{equation}
which can be positive, null, or~negative.

The third case corresponds to $\omega = M$, for~which $r_{H}=2M$. Therefore, analogously to the other cases, we can compute the following:
\begin{subequations}
\begin{align}
        T^{^{(A)}}&=T_0,\\
        S^{^{(A)}}&=4\pi M^2,\\
        C_V^{^{(A)}}&=C_0.
    \end{align}
\end{subequations}

Finally, the~free energy is
\begin{equation}
    F^{^{(A)}}=\frac{M}{2},
\end{equation}
which is minimized only for $M=0$; thus, the system is globally~unstable.

This case reproduces the behavior of general relativity's non-rotating black holes. Since the area and quantum tunneling methods lead to different results in this second case as well, the first law must be revisited~\cite{Gomes:2020bhsthermodynamics}. However, in~this case, it is not possible to fully determine the function $L^{^{(A)}}$ since the quantum tunneling leads to divergent~quantities.\\

\section{Reissner--Nordstrøm-Like Black~Hole}
\label{Sec:RN}

In this section, we will analyze black hole solutions of the form of Reissner--Nordstrøm, i.e.,~black holes which are static and have mass and electric charge. 
In order to describe the gravitational field outside a charged, non-rotating, spherically symmetric body, we consider the electrostatic four-potential $\Phi_\mu=(-\phi(r),0,0,0)$, with~$\phi(r)$ as the scalar potential. Using the relation (\ref{Tensor_Eletro}), it is possible to see that $\mathcal{L}^{^{(EM)}}=\frac{1}{2} e^{-\alpha(r)-\nu(r)} \phi'(r)^2$, and the energy momentum-tensor is such that
\begin{equation}
    T{^{^{(EM)}}}^\mu_\nu=\text{diag}(-1,-1,1,1)\frac{1}{2}e^{-\alpha(r)-\beta(r)}\phi'(r)^2.
\end{equation}

Applying the previous result into the Maxwell Equation~(\ref{Maxwell_eqs}), it follows that
\begin{equation}
    \phi'(r)=-e^{\frac{1}{2}(\alpha(r)+\beta(r))} \frac{c_1}{f_2(\bar{R}) r^2},
    \label{eq-dphi}
\end{equation}
with $c_1$ being some constant, which we identify with the electric charge, $Q$, in~analogy with the general relativity~result.

Analogously to the Schwarzchild-like solutions, we shall look into vacuum and the first simplest contribution from non-vacuum Lagrangian~density.

\subsection{Vacuum~Background}

We now consider a vacuum background. Thus, the~contribution to the energy-momentum tensor is given only by the black hole geometry seen from the infinity as a point~charge.

First of all, it should be noted that $\Theta(\bar{R})\neq 0$. Otherwise, Equation~(\ref{trace-free_eqs}) implies zero energy-momentum tensor, since $T^{^{(EM)}}=0$. Therefore, the~only applicable Weyl vector is the radial one, Equation~(\ref{eqn:ansatz1}). 

From the trace of the field Equation~(\ref{trace}), we have
\begin{equation}
    f_1(\bar{R})=\frac{1}{2}\bar{R} \Theta(\bar{R}). 
\end{equation}

Applying this result, together with the fact that $F_1(\bar{R})=\frac{d f_1(\bar{R})}{d \bar{R}}$ and the constraint (\ref{eqconstraint}), we obtain
\begin{equation}
    F_1(\bar{R})=\frac{1}{2}\left(1-A(r) \frac{\bar{R}}{\bar{R}'} \right)\Theta(\bar{R}),
\end{equation}
when $\bar{R}'\neq 0$.

Using this relation, together with $\Theta(\bar{R})=F_1(\bar{R})+\mathcal{L}F_2(\bar{R})$ and the non-conservation equation to energy-momentum tensor (\ref{nonconserveq}), it is possible to observe that $f_2(\bar{R})=0$. Thus, we conclude that it is not possible to derive a charged black hole solution, in~a vacuum, for~any static and spherical metric in the model under~study. 

If $\bar{R}'= 0$, it is also impossible to obtain a black hole solution. In~fact, $\bar{R}'= 0$ implies that the curvature scalar is constant, so $\Theta(\bar{R})$ is also constant. From~the constraint (\ref{eqconstraint}), this implies  that $A(r)=0$. 

\subsection{Cosmological Constant~Background}

We now consider a charged black hole immersed in a cosmological constant background. Then, the~Lagrangian density is given by $\mathcal{L}=\mathcal{L}^{^{(EM)}}+\mathcal{L}^{^{(\Lambda)}}$ and the components of the energy momentum tensor are $T_{\mu\nu}=T^{^{(EM)}}_{\mu\nu}+T^{^{(\Lambda)}}_{\mu\nu}$, as described~above.

It should be noted that, again, $\Theta(\bar{R})\neq 0$. Otherwise, Equation~(\ref{trace-free_eqs}) implies zero energy-momentum tensor, since $T^{^{(EM)}}=0$. Therefore, the~only viable ansatz for the Weyl vector is the radial one, Equation~(\ref{eqn:ansatz1}). 

By combining all trace-free Equation~(\ref{trace-free_eqs}), it is possible to find a similar equation to (\ref{eq25}). Therefore, we will use the same relation $\beta(r)=-\alpha(r)+\epsilon$. With~this assumption, it is possible to see that the solution to $A(r)$ is, again,
\begin{equation}
    A(r)=\frac{2}{r+\omega},
\end{equation}
with $\omega>0$. 

Thus, from~(\ref{eqconstraint}), we obtain
\begin{eqnarray}
    \Theta(\bar{R})=\frac{\xi}{(r+\omega)^2},
\end{eqnarray}
with $\xi$ some integration constant that we will define~later.

Using this result in all previous equations, it is possible to obtain that the appropriated model is given by
\begin{subequations} \label{Model-RN-Lambda}
\begin{align}
    f_1(\bar{R})&=\gamma \bar{R}^2+2\Lambda \zeta\\
    f_2(\bar{R})&=\zeta,
    \end{align}
\end{subequations}
where $\gamma=\xi \frac{(6\tilde{Q}^2+6M\omega+\omega^2)}{72 \tilde{Q}^2}$, $\xi=\frac{\omega^2}{4}\zeta$, $\zeta$ is some constant, and $\tilde{Q}$ is the dressed charge such that Equation~(\ref{eq-dphi}) gives
\begin{equation}
    \phi(r)=\frac{\tilde{Q}}{r},
\end{equation}
and we have the relation
\begin{equation}
    Q^2=\zeta^2 \tilde{Q}^2 \left( 1+6\left(\frac{M}{\omega}+\frac{\tilde{Q}^2}{\omega^2}\right) \right)^{-1},
\end{equation}
where $Q$ is the usual~charge. 

The metric solution to this problem is given by
\begin{subequations}\label{solutionRN}
\begin{align}
        e^{\alpha(r)}&=1-\frac{2M}{r}+\frac{\tilde{Q}^2}{r^2}+\frac{2\left(\omega^2+3M\omega+2\tilde{Q}^2\right)}{\omega^2} \frac{r}{\omega}+\frac{4M+\omega}{\omega}\left( \frac{r}{\omega} \right)^2,\\
        e^{\beta(r)}&=\frac{1+6\left(\frac{M}{\omega}+\frac{\tilde{Q}^2}{\omega^2}\right)}{1-\frac{2M}{r}+\frac{\tilde{Q}^2}{r^2}+\frac{2\left(\omega^2+3M\omega+2\tilde{Q}^2\right)}{\omega^2} \frac{r}{\omega}+\frac{4M+\omega}{\omega}\left( \frac{r}{\omega} \right)^2}.
\end{align}
\end{subequations}

The curvature scalar takes the form
\begin{equation}
    \bar{R}=\frac{36 \tilde{Q}^2}{\left( \omega^2+6M\omega+6\tilde{Q}^2 \right)\left(r+\omega\right)^2}.
    \label{RN-curvature-scalar}
\end{equation}

It is easy to see that when considering the limit $\tilde{Q}\rightarrow0$, the~curvature scalar $\bar{R}\rightarrow0$ and the solution (\ref{solutionRN}) converge to a Schwarzschild-like solution (\ref{solution-radialSchw}). 

\begin{figure}[H]
	\includegraphics[width=0.8\textwidth]{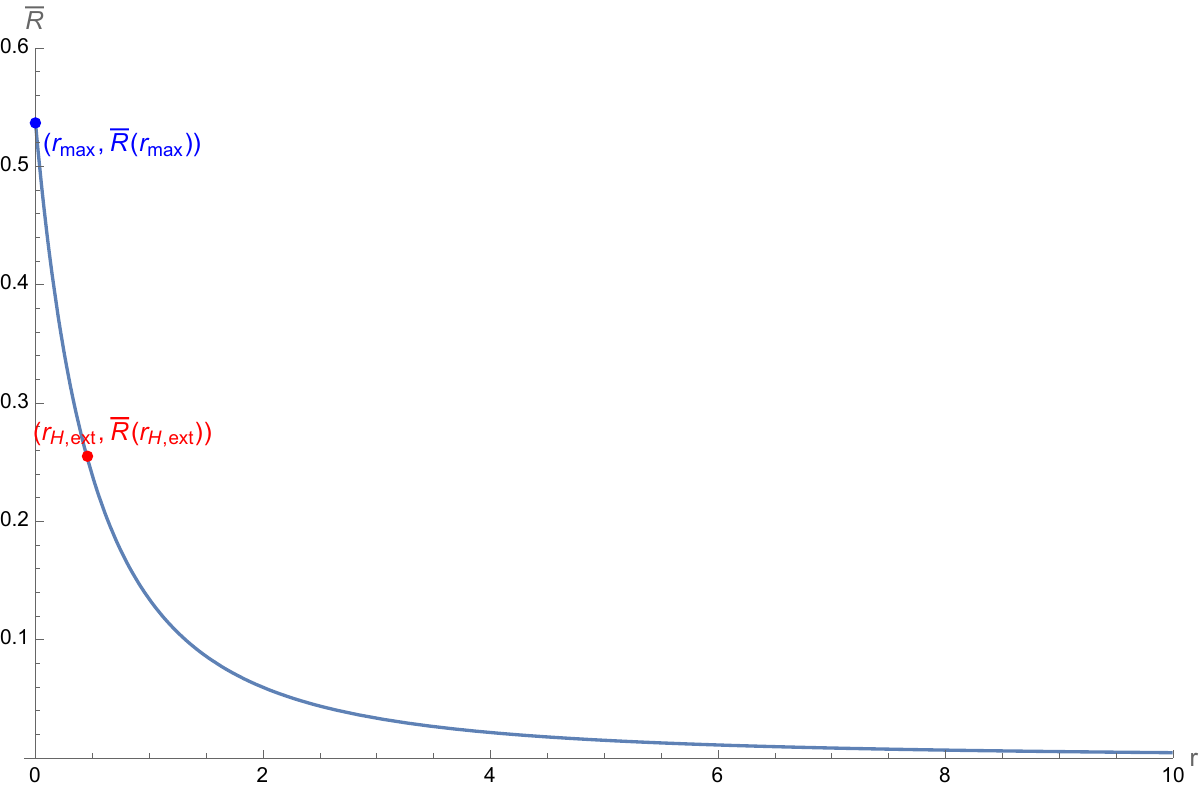} \vspace{-8pt}
	\caption{Global behavior of the scalar curvature $\bar{R}$ as a function of the distance, assuming $M=10$, $\omega=1$, and $\tilde{Q}=1$.}\label{CASE-RN-Scalar}
\end{figure}

Analyzing, globally, the~curvature scalar (\ref{RN-curvature-scalar}) is positive for all $r$ and it decreases with distance from the black hole. The~maximum value occurs when $r_{\text{max}}=0$, taking the value $\bar{R}_{\text{max}}=\frac{36 \tilde{Q}^2}{\omega^2 (\omega^2+6M\omega+6\tilde{Q}^2)}$. In~Figure~\ref{CASE-RN-Scalar}, we represent the global behavior of the scalar curvature as a function of the distance, considering a specific case, where the point at which the external event horizon occurs is~represented.
\vspace{0.5cm}

\vspace{0.5cm}

The Ricci scalar takes the form
\begin{equation}
    R=-\frac{12\left( (4M+\omega)r^2+ (\omega^2+3M\omega+2\tilde{Q}^2)r-(\tilde{Q}^2+M \omega)\omega \right)}{r^2 \omega \left( \omega^2+6M\omega+2\tilde{Q}^2 \right)}.
    \label{RN-Ricci}
\end{equation}

\begin{figure}[H]
	\includegraphics[width=0.8\textwidth]{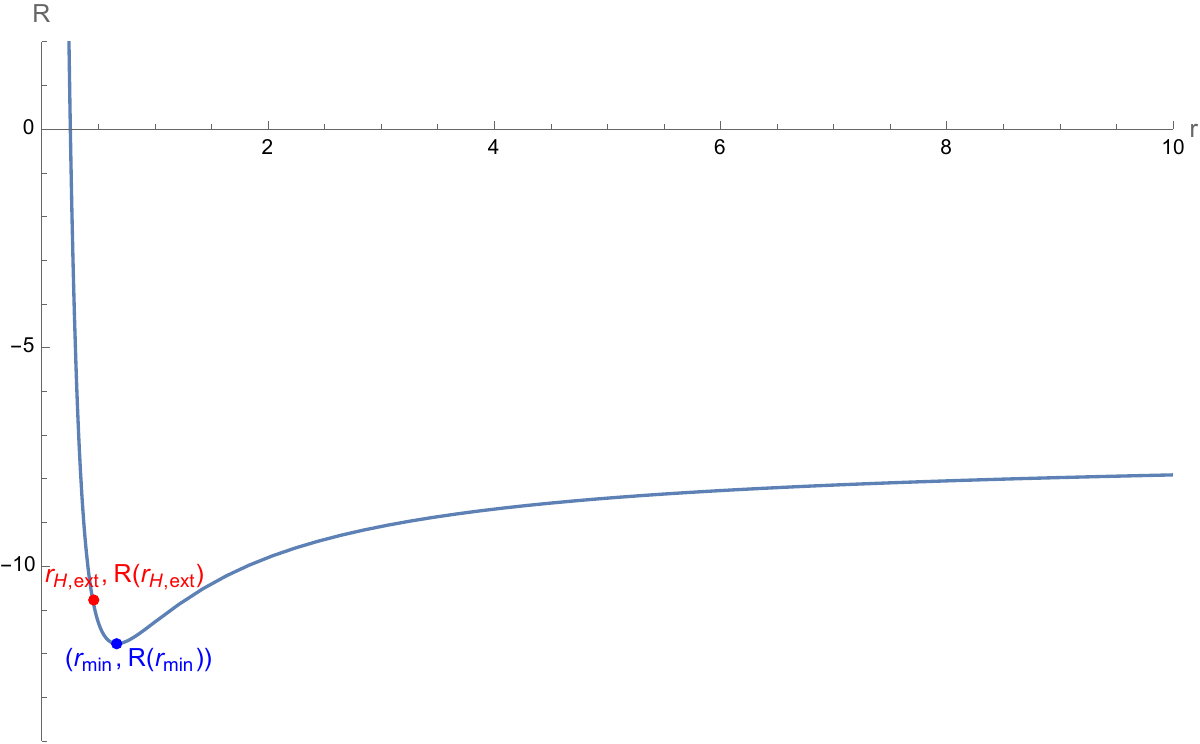}  \vspace{-8pt}
	\caption{Global behavior of the standard Ricci scalar (built from the Levi--Civita part of the connection) as a function of the distance, assuming $M=10$, $\omega=1$, and $\tilde{Q}=1$.}\label{CASE-RN-Ricci}
\end{figure}

It is easy to see that when considering the limit $\tilde{Q}\rightarrow0$, the~Ricci scalar $R$ converges to a Schwarzschild-like solution (\ref{Ricci-sol-CASE1}). In~Figure~\ref{CASE-RN-Ricci}, we represent the global behavior of the scalar curvature as a function of the distance, considering a specific case, where the point at which the external event horizon occurs is~represented. 

Analyzing, globally, the~Ricci scalar (\ref{RN-Ricci}), it is possible to see that when $r\to 0$, the~Ricci scalar goes to infinity, $R\to +\infty$. When $r\to\infty$, the~Ricci scalar converges to a non-null constant, $R\to -\frac{12(\omega+4M)}{\omega(\omega^2+6M\omega+6\tilde{Q}^2)}$. We can also note that the minimum value of Ricci scalar occurs when $r_{\text{min}}=\frac{2\omega \left(\tilde{Q}^2+M\omega\right)}{\omega^2+3M\omega+2\tilde{Q}^2}$ and takes the value $R_{\text{min}}=-\frac{3\left( 4\tilde{Q}^2(\tilde{Q}^2+7M\omega+2\omega^2)+\omega^2(5M+\omega)^2 \right)}{\omega^2 (\tilde{Q}^2+M\omega)(\omega^2+6M\omega+6\tilde{Q}^2)}$.\\

In order to assess the nature of the singularities in this result, we compute the Kretschmann invariant. Thus, for~this case, we obtain
\begin{equation}
    K= \frac{h(r)}
{r^8 (r + w)^4 \left(6 Q^2 + w (6 M + w)\right)^2},
    \label{Kretschmann-BH2}
\end{equation}
where $h(r)$ is is a large polynomial expression of degree 8, which we opt to not display as it is not relevant for this analysis. It is possible to see that the Kretschmann invariant only diverges when $r=0$ for any $M>0$ and $\omega>0$. Therefore, $r=0$ is an essential singularity in~the center of the black~hole.\\

Due to the complexity of the solution obtained, it is not possible to provide a generic and complete analysis. We know that, under~certain circumstances, we are in the presence of one or two event horizons, but~it is difficult to rigorously define these~values.

Tables~\ref{Table-RN-BH-1}, \ref{Table-RN-BH-2}, and \ref{Table-RN-BH-3} represent the estimated values for the internal, $r_{H,\text{int}}$, and~external, $r_{H,\text{ext}}$, event horizon considering the mass values $10^{-2}$, 1, and $10^{2}$, respectively. We can conclude that, for~the cases where there   {are} event horizons which correspond to lower values for the effective charge, provided the Weyl constant is above a certain lower limit, the~values of both horizons are strikingly~similar.

\begin{table}[H]
	\centering
	\footnotesize
	\begin{tabular}{|c|c|c|c|}
		\hline
		$\tilde{Q}$      & $\omega$        & $r_{H,\text{int}}$  & $r_{H,\text{ext}}$  \\ \specialrule{1.5pt}{-1pt}{-1pt }
		\multirow{9}{*}{$10^{-4}$} & $10^{-7}$ & no $r_{H,\text{int}}$          & no $r_{H,\text{ext}}$          \\ \cline{2-4}
		& $10^{-6}$ & no $r_{H,\text{int}}$          & no $r_{H,\text{ext}}$          \\ \cline{2-4}
		& $10^{-5}$ & $5.0425 \times 10^{-7}$ & $4.6474 \times 10^{-6}$ \\ \cline{2-4}
		& $10^{-4}$ & $5.0005 \times 10^{-7}$ & $4.9541 \times 10^{-5}$ \\ \cline{2-4}
		& $10^{-3}$ & $5.0001 \times 10^{-7}$ & $4.8747 \times 10^{-4}$ \\ \cline{2-4}
		& $10^{-2}$ & $5.0001 \times 10^{-7}$ & $3.9997 \times 10^{-3}$ \\ \cline{2-4}
		& $10^{-1}$ & $5.0001 \times 10^{-7}$ & $1.4285 \times 10^{-2}$ \\ \cline{2-4}
		& $10^{1}$  & $5.0001 \times 10^{-7}$ & $1.9920 \times 10^{-2}$ \\ \cline{2-4}
		& $10^{2}$  & $5.0001 \times 10^{-7}$ & $1.9992 \times 10^{-2}$ \\ \specialrule{1.5pt}{-1pt}{-1pt }
		\multirow{9}{*}{$10^{-3}$} & $10^{-5}$ & no $r_{H,\text{int}}$          & no $r_{H,\text{ext}}$          \\ \cline{2-4}
		& $10^{-4}$ & no $r_{H,\text{int}}$          & no $r_{H,\text{ext}}$          \\ \cline{2-4}
		& $10^{-3}$ & $5.0568 \times 10^{-5}$ & $4.5281 \times 10^{-4}$ \\ \cline{2-4}
		& $10^{-2}$ & $5.0131 \times 10^{-5}$ & $3.9677 \times 10^{-3}$ \\ \cline{2-4}
		& $10^{-1}$ & $5.0126 \times 10^{-5}$ & $1.4247 \times 10^{-2}$ \\ \cline{2-4}
		& $10^{0}$  & $5.0126 \times 10^{-5}$ & $1.9183 \times 10^{-2}$ \\ \cline{2-4}
		& $10^{1}$  & $5.0126 \times 10^{-5}$ & $1.9870 \times 10^{-2}$ \\ \cline{2-4}
		& $10^{3}$  & $5.0126 \times 10^{-5}$ & $1.9949 \times 10^{-2}$ \\ \cline{2-4}
		& $10^{5}$  & $5.0126 \times 10^{-5}$ & $1.9950 \times 10^{-2}$ \\ \specialrule{1.5pt}{-1pt}{-1pt }
		$10^{-2}$      & for all $w>0$       & no $r_{H,\text{int}}$  & no $r_{H,\text{ext}}$  \\ \specialrule{1.5pt}{-1pt}{-1pt }
		$10^{-1}$      & for all $w>0$        & no $r_{H,\text{int}}$  & no $r_{H,\text{ext}}$  \\ \specialrule{1.5pt}{-1pt}{-1pt }
		$1$      & for all $w>0$        & no $r_{H,\text{int}}$  & no $r_{H,\text{ext}}$  \\ \specialrule{1.5pt}{-1pt}{-1pt }
		$10^1$      & for all $w>0$        & no $r_{H,\text{int}}$  & no $r_{H,\text{ext}}$  \\ \specialrule{1.5pt}{-1pt}{-1pt }
		$10^2$      & for all $w>0$        & no $r_{H,\text{int}}$  & no $r_{H,\text{ext}}$  \\ \specialrule{1.5pt}{-1pt}{-1pt }
		$10^3$      & for all $w>0$        & no $r_{H,\text{int}}$  & no $r_{H,\text{ext}}$  \\ \specialrule{1.5pt}{-1pt}{-1pt }
		$10^4$      & for all $w>0$        & no $r_{H,\text{int}}$  & no $r_{H,\text{ext}}$  \\ \hline
	\end{tabular}
	\caption{Estimated values for the internal, $r_{H,\text{int}}$, and~external, $r_{H,\text{ext}}$, event horizon considering the mass value $M=10^{-2}$ and different values for the charge and the Weyl parameters, $\tilde{Q}$ and $\omega$, respectively.}
	\label{Table-RN-BH-1}
\end{table}

\hspace{-0.5cm}
\begin{minipage}{0.5\textwidth}
\begin{table}[H]
	\centering
	\scriptsize
	\begin{tabular}{|c|c|c|c|}
		\hline
		$\tilde{Q}$      & $\omega$        & $r_{H,\text{int}}$  & $r_{H,\text{ext}}$  \\ \specialrule{1.5pt}{0pt}{0pt}
		\multirow{12}{*}{$10^{-4}$} & $10^{-9}$ & no $r_{H,\text{int}}$ & no $r_{H,\text{ext}}$ \\ \cline{2-4}
		& $10^{-8}$ & no $r_{H,\text{int}}$ & no $r_{H,\text{ext}}$ \\ \cline{2-4}
		& $10^{-7}$ & $5.0423 \times 10^{-9}$ & $4.6487 \times 10^{-8}$ \\ \cline{2-4}
		& $10^{-6}$ & $5.0004 \times 10^{-9}$ & $ 4.9665 \times 10^{-7}$ \\ \cline{2-4}
		& $10^{-5}$ & $5.0000 \times 10^{-9}$ & $ 4.9967 \times 10^{-6}$ \\ \cline{2-4}
		& $10^{-4}$ & $5.0000 \times 10^{-9}$ & $ 4.9995 \times 10^{-5}$ \\ \cline{2-4}
		& $10^{-3}$ & $5.0000 \times 10^{-9}$ & $ 4.9987 \times 10^{-4}$ \\ \cline{2-4}
		& $10^{-2}$  & $5.0000 \times 10^{-9}$ & $ 4.9875 \times 10^{-3}$ \\ \cline{2-4}
		& $10^{-1}$  & $5.0000 \times 10^{-9}$ & $ 4.8780 \times 10^{-2}$ \\ \cline{2-4}
		& $1$  & $5.0000 \times 10^{-9}$ & $ 4.0000 \times 10^{-1}$ \\ \cline{2-4}
		& $10^{1}$  & $5.0000 \times 10^{-9}$ & $ 1.4286 \times 10^{0}$ \\ \cline{2-4}
		& $10^{2}$  & $5.0000 \times 10^{-9}$ & $ 1.9231 \times 10^{0}$ \\ \cline{2-4}
		& $10^{4}$  & $5.0000 \times 10^{-9}$ & $ 1.9992 \times 10^{0}$ \\ \specialrule{1.5pt}{0pt}{0pt}
		\multirow{9}{*}{$10^{-3}$} & $10^{-7}$ & no $r_{H,\text{int}}$ & no $r_{H,\text{ext}}$ \\ \cline{2-4}
		& $10^{-6}$ & no $r_{H,\text{int}}$ & no $r_{H,\text{ext}}$ \\ \cline{2-4}
		& $10^{-5}$ & $ 5.0423 \times 10^{-7}$ & $  4.6487 \times 10^{-6}$ \\ \cline{2-4}
		& $10^{-4}$ & $ 5.0004 \times 10^{-7}$ & $  4.9664 \times 10^{-5}$ \\ \cline{2-4}
		& $10^{-3}$ & $ 5.0000 \times 10^{-7}$ & $  4.9954\times 10^{-4}$ \\ \cline{2-4}
		& $10^{-2}$  & $ 5.0000 \times 10^{-7}$ & $  4.9872 \times 10^{-3}$ \\ \cline{2-4}
		& $10^{-1}$  & $ 5.0000 \times 10^{-7}$ & $ 4.8780 \times 10^{-2}$ \\ \cline{2-4}
		& $1$  & $ 5.0000 \times 10^{-7}$ & $ 4.0000 \times 10^{-1}$ \\ \cline{2-4}
		& $10^{1}$  & $ 5.0000 \times 10^{-7}$ & $ 1.4286 \times 10^{0}$ \\ \specialrule{1.5pt}{0pt}{0pt}
		\multirow{9}{*}{$10^{-2}$} & $10^{-5}$ & no $r_{H,\text{int}}$ & no $r_{H,\text{ext}}$ \\ \cline{2-4}
		& $10^{-4}$ & no $r_{H,\text{int}}$ & no $r_{H,\text{ext}}$ \\ \cline{2-4}
		& $10^{-3}$ & $ 5.0425 \times 10^{-5}$ & $ 4.6474 \times 10^{-4}$ \\ \cline{2-4}
		& $10^{-2}$ & $ 5.0005 \times 10^{-5}$ & $  4.9541 \times 10^{-3}$ \\ \cline{2-4}
		& $10^{-1}$ & $ 5.0001 \times 10^{-5}$ & $ 4.8747 \times 10^{-2}$ \\ \cline{2-4}
		& $1$  & $ 5.0001 \times 10^{-5}$ & $   3.9997 \times 10^{-1}$ \\ \cline{2-4}
		& $10^{1}$  & $ 5.0001 \times 10^{-5}$ & $ 1.4285 \times 10^{0}$ \\ \cline{2-4}
		& $10^{2}$  & $ 5.0001 \times 10^{-5}$ & $ 1.9230 \times 10^{0}$ \\ \cline{2-4}
		& $10^{3}$  & $ 5.0001 \times 10^{-5}$ & $ 1.9920 \times 10^{0}$ \\ \specialrule{1.5pt}{0pt}{0pt}
		\multirow{9}{*}{$10^{-1}$} & $10^{-3}$ & no $r_{H,\text{int}}$ & no $r_{H,\text{ext}}$ \\ \cline{2-4}
		& $10^{-2}$ & no $r_{H,\text{int}}$ & no $r_{H,\text{ext}}$ \\ \cline{2-4}
		& $10^{-1}$ & $ 5.0568 \times 10^{-3}$ & $ 4.5281 \times 10^{-2}$ \\ \cline{2-4}
		& $1$ & $ 5.0131 \times 10^{-3}$ & $  3.9677 \times 10^{-1}$ \\ \cline{2-4}
		& $10^{1}$ & $ 5.0126 \times 10^{-3}$ & $  1.4247 \times 10^{0}$ \\ \cline{2-4}
		& $10^{2}$  & $ 5.0126 \times 10^{-3}$ & $  1.9183 \times 10^{0}$ \\ \cline{2-4}
		& $10^{3}$  & $ 5.0126 \times 10^{-3}$ & $  1.9870 \times 10^{0}$ \\ \cline{2-4}
		& $10^{4}$  & $ 5.0126 \times 10^{-3}$ & $ 1.9942 \times 10^{0}$ \\ \cline{2-4}
		& $10^{5}$  & $ 5.0126 \times 10^{-3}$ & $ 1.9949 \times 10^{0}$ \\ \specialrule{1.5pt}{0pt}{0pt}
		$1$      & for all $w>0$       & no $r_{H,\text{int}}$ & no $r_{H,\text{ext}}$ \\ \specialrule{1.5pt}{0pt}{0pt}
		$10^1$      & for all $w>0$        & no $r_{H,\text{int}}$ & no $r_{H,\text{ext}}$ \\ \specialrule{1.5pt}{0pt}{0pt}
		$10^2$      & for all $w>0$        & no $r_{H,\text{int}}$ & no $r_{H,\text{ext}}$ \\ \specialrule{1.5pt}{0pt}{0pt}
		$10^3$      & for all $w>0$        & no $r_{H,\text{int}}$ & no $r_{H,\text{ext}}$ \\ \specialrule{1.5pt}{0pt}{0pt}
		$10^4$      & for all $w>0$        & no $r_{H,\text{int}}$ & no $r_{H,\text{ext}}$ \\ \hline
	\end{tabular}
	\caption{Estimated values for the internal, $r_{H,\text{int}}$, and external, $r_{H,\text{ext}}$, event horizon considering the mass value $M=1$ and different values for the charge and the Weyl parameters, $\tilde{Q}$ and $\omega$, respectively.}
	\label{Table-RN-BH-2}
\end{table}

\end{minipage}
\hspace{0.5cm} 
\begin{minipage}{0.5\textwidth}
\begin{table}[H]
	\centering
	\scriptsize
	\begin{tabular}{|c|c|c|c|}
		\hline
		$\tilde{Q}$ & $\omega$ & $r_{H,\text{int}}$ & $r_{H,\text{ext}}$ \\
		\specialrule{1.5pt}{0pt}{0pt}
		
		\multirow{12}{*}{$10^{-4}$} & $10^{-11}$ & no $r_{H,\text{int}}$ & no $r_{H,\text{ext}}$ \\ \cline{2-4}
		& $10^{-10}$ & no $r_{H,\text{int}}$ & no $r_{H,\text{ext}}$ \\ \cline{2-4}
		& $10^{-9}$  & $5.0423 \times 10^{-11}$ & $4.6487 \times 10^{-10}$ \\ \cline{2-4}
		& $10^{-8}$  & $5.0004 \times 10^{-11}$ & $4.9665 \times 10^{-9}$ \\ \cline{2-4}
		& $10^{-7}$  & $5.0000 \times 10^{-11}$ & $4.9967 \times 10^{-8}$ \\ \cline{2-4}
		& $10^{-6}$  & $5.0000 \times 10^{-11}$ & $4.9997 \times 10^{-7}$ \\ \cline{2-4}
		& $10^{-4}$  & $5.0000 \times 10^{-11}$ & $5.0000 \times 10^{-5}$ \\ \cline{2-4}
		& $10^{-2}$  & $5.0000 \times 10^{-11}$ & $4.9999 \times 10^{-3}$ \\ \cline{2-4}
		& $1$        & $5.0000 \times 10^{-11}$ & $4.9875 \times 10^{-1}$ \\ \cline{2-4}
		& $10^{1}$   & $5.0000 \times 10^{-11}$ & $4.8780 \times 10^{0}$ \\ \cline{2-4}
		& $10^{2}$   & $5.0000 \times 10^{-11}$ & $4.0000 \times 10^{1}$ \\ \cline{2-4}
		& $10^{3}$   & $5.0000 \times 10^{-11}$ & $1.4286 \times 10^{2}$ \\ \cline{2-4}
		& $10^{4}$   & $5.0000 \times 10^{-11}$ & $1.9231 \times 10^{2}$ \\ \cline{2-4}
		& $10^{6}$   & $5.0000 \times 10^{-11}$ & $1.9992 \times 10^{2}$ \\ 
		\specialrule{1.5pt}{0pt}{0pt}
		
		\multirow{12}{*}{$10^{-3}$} & $10^{-9}$  & no $r_{H,\text{int}}$ & no $r_{H,\text{ext}}$ \\ \cline{2-4}
		& $10^{-8}$  & no $r_{H,\text{int}}$ & no $r_{H,\text{ext}}$ \\ \cline{2-4}
		& $10^{-7}$  & $5.0423 \times 10^{-9}$ & $4.6487 \times 10^{-8}$ \\ \cline{2-4}
		& $10^{-6}$  & $5.0004 \times 10^{-9}$ & $4.9665 \times 10^{-7}$ \\ \cline{2-4}
		& $10^{-5}$  & $5.0000 \times 10^{-9}$ & $4.9967 \times 10^{-6}$ \\ \cline{2-4}
		& $10^{-4}$  & $5.0000 \times 10^{-9}$ & $4.9997 \times 10^{-5}$ \\ \cline{2-4}
		& $10^{-3}$  & $5.0000 \times 10^{-9}$ & $5.0000 \times 10^{-4}$ \\ \cline{2-4}
		& $10^{-2}$  & $5.0000 \times 10^{-9}$ & $4.9999 \times 10^{-3}$ \\ \cline{2-4}
		& $10^{-1}$  & $5.0000 \times 10^{-9}$ & $4.9987 \times 10^{-2}$ \\ \cline{2-4}
		& $1$        & $5.0000 \times 10^{-9}$ & $4.9875 \times 10^{-1}$ \\ \cline{2-4}
		& $10^{1}$   & $5.0000 \times 10^{-9}$ & $4.8780 \times 10^{0}$ \\ \cline{2-4}
		& $10^{2}$   & $5.0000 \times 10^{-9}$ & $4.0000 \times 10^{1}$ \\ \cline{2-4}
		& $10^{3}$   & $5.0000 \times 10^{-9}$ & $1.4286 \times 10^{2}$ \\ \cline{2-4}
		& $10^{5}$   & $5.0000 \times 10^{-9}$ & $1.9920 \times 10^{2}$ \\ 
		\specialrule{1.5pt}{0pt}{0pt}
		
		\multirow{12}{*}{$10^{-2}$} & $10^{-7}$  & no $r_{H,\text{int}}$ & no $r_{H,\text{ext}}$ \\ \cline{2-4}
		& $10^{-6}$  & no $r_{H,\text{int}}$ & no $r_{H,\text{ext}}$ \\ \cline{2-4}
		& $10^{-5}$  & $5.0423 \times 10^{-7}$ & $4.6487 \times 10^{-6}$ \\ \cline{2-4}
		& $10^{-4}$  & $5.0004 \times 10^{-7}$ & $4.9665 \times 10^{-5}$ \\ \cline{2-4}
		& $10^{-3}$  & $5.0000 \times 10^{-7}$ & $4.9967 \times 10^{-4}$ \\ \cline{2-4}
		& $10^{-2}$  & $5.0000 \times 10^{-7}$ & $4.9995 \times 10^{-3}$ \\ \cline{2-4}
		& $10^{-1}$  & $5.0000 \times 10^{-7}$ & $4.9987 \times 10^{-2}$ \\ \cline{2-4}
		& $1$        & $5.0000 \times 10^{-7}$ & $4.9875 \times 10^{-1}$ \\ \cline{2-4}
		& $10^{1}$   & $5.0000 \times 10^{-7}$ & $4.8780 \times 10^{0}$ \\ \cline{2-4}
		& $10^{2}$   & $5.0000 \times 10^{-7}$ & $4.0000 \times 10^{1}$ \\ \cline{2-4}
		& $10^{3}$   & $5.0000 \times 10^{-7}$ & $1.4286 \times 10^{2}$ \\ \cline{2-4}
		& $10^{6}$   & $5.0000 \times 10^{-7}$ & $1.9992 \times 10^{2}$ \\
		\specialrule{1.5pt}{0pt}{0pt}
		
		$10^{-1}$ & for all $w > 0$ & no $r_{H,\text{int}}$ & no $r_{H,\text{ext}}$ \\
		\specialrule{1.5pt}{0pt}{0pt}
		
		$1$ & for all $w > 0$ & no $r_{H,\text{int}}$ & no $r_{H,\text{ext}}$ \\
		\specialrule{1.5pt}{0pt}{0pt}
		
		$10^{1}$ & for all $w > 0$ & no $r_{H,\text{int}}$ & no $r_{H,\text{ext}}$ \\
		\specialrule{1.5pt}{0pt}{0pt}
		
		$10^{2}$ & for all $w > 0$ & no $r_{H,\text{int}}$ & no $r_{H,\text{ext}}$ \\
		\specialrule{1.5pt}{0pt}{0pt}
		
		$10^{3}$ & for all $w > 0$ & no $r_{H,\text{int}}$ & no $r_{H,\text{ext}}$ \\
		\specialrule{1.5pt}{0pt}{0pt}
		
		$10^{4}$ & for all $w > 0$ & no $r_{H,\text{int}}$ & no $r_{H,\text{ext}}$ \\
		\hline
	\end{tabular}
\caption{Estimated values for the internal, $r_{H,\text{int}}$, and external, $r_{H,\text{ext}}$, event horizon considering the mass value $M=10^2$ and different values for the charge and the Weyl parameters, $\tilde{Q}$ and $\omega$, respectively.}
\label{Table-RN-BH-3}
\end{table}
\end{minipage}
\vspace{0.5cm}

Since the analytical solutions are too cumbersome to be treated in the derivation of the corresponding thermodynamical quantities, this analysis will not be pursued for the Reissner--Nordstrøm~case.

\section{Conclusions}
\label{Sec:conclusions}

In this work, we analyzed black hole solutions of a non-minimally coupled Weyl connection gravity in the form of generalized Schwarzschild and Reissner--Nordstrøm solutions both in a vacuum and in the presence of matter fluids behaving as a cosmological constant. The~Weyl connection is a particular case of non-metricity, and, in the context of these alternative theories of gravity, leads to black hole solutions for which the scalar curvature is~non-vanishing.

When in a vacuum, the~model under study is equivalent to $f(R)$ theories with the Weyl connection. Thus, in~comparison with constant curvature solutions for vacuum metric $f(R)$ theories, our model introduces a term linear to the radius, $r$, and~a corrected cosmological constant in the solutions for the metric functions for Schwarzschild-like black holes. In~addition, no solutions of the Reissner--Nordstrøm form are~found.

However, the~non-minimally coupled Weyl connection gravity model for non-vacuum solutions behaves differently. In~particular, matter fields of the form of a cosmological constant lead to Schwarzschild-like black holes exhibiting a behavior analogous to the vacuum case with a contribution also arising from this matter Lagrangian choice. This is mathematically equivalent to a reparametrization of vacuum $f(R)$ theories; however, physically, they are different because one is assuming a contribution from vacuum energy or matter fields behaving as a cosmological constant that is different from an integration constant or a numerical rescaling of functions, in~the same spirit of Ref.~\cite{Bertolami:2015bhs}. Moreover, Reissner--Nordstrøm solutions can be found with a linear term in $r$ and corrected/dressed charge and cosmological constant in the solution for the metric~functions. 

There are several studies considering $f(R)$ theories that analyze black hole solutions. In~Ref.~\cite{Capozziello:2010bhs}, the~authors present a Schwarzchild-like solution with an asymptotic behavior similar to ours; the~universe expands with a $r^2$ factor. However, in~our case, a linear contribution also appears. Additionally, in Ref.~\cite{CruzDombriz:2009bhsfR}, it is possible to find, in~$f(R)$ theory, a~solution to a charged black hole in an expansive anti-de Sitter space with $r^2$. Therefore, our Reissner--Nordstrøm solution presents the same asymptotic behavior; however, also with an additional linear~term.

Similar to Ref.~\cite{Harko:2022}, where the authors provided numerical black hole solutions in the Weyl conformal geometry that also have an expansion behavior, our study found an exact solution that presents a Schwarzschild term inversely proportional to r and two more terms proportional to $r$ and $r^2$. Moreover, black hole solutions in 4D Einstein–Gauss–Bonnet theory exhibit, like ours, an~asymptotically non-flat behavior given the $r^2$ factor in the metric solution~\cite{Glavan2020}.

In fact, our analysis shows that the $r=0$ singularity is an essential one; thus, we are not able to remove it. Since the Kretschmann invariants for the cases we analyzed depend only on the radial coordinate and on constants, we cannot find a relation among the numerator components such that we could eliminate the exact dependency on $r$ at the denominator. However, if~we consider a Hayward metric ansatz, we may circumvent this issue, thus having a model lacking a central singularity. This leads to an entire new work in the~future.

Other solutions of black holes can exist in this gravity model. In~particular, Kerr solutions, even in the form of slowly-rotating spherically symmetric space--times or black hole solutions different from analytics ones from general relativity, may be obtained in future work, but~they fall outside of the scope of the present paper. Moreover, the~results found in this work may be relevant to discriminate between modified gravity theories, provided that numerical simulations allow the incorporation of extra degrees of freedom. In~particular, the~gravitational wave data from the collision of black holes may be an interesting~avenue of study.

\vspace{6pt} 

\section*{Acknowledgements}
M. M. Lima acknowledges support from Fundo Regional da Ciência e Tecnologia and Azores Government through the Fellowship M3.1.a/F/001/2022.
\appendix
\section[\appendixname~\thesection]{Generalized Riemann Curvature Tensor Components}\label{sec:annex}

In this section, we will enumerate the non-vanished generalized Riemann curvature tensor components given by (\ref{Riemann-total}). Note that the tensor is antisymmetric in the last two indices, $\bar{R}^\rho_{\mu\sigma\nu}=-\bar{R}^\rho_{\mu\nu\sigma}$. We will present only one of~them.\\

\subsection[\appendixname~\thesection]{First Ansatz: $A_\mu=(0,A(r),0,0)$}
 
 Considering (\ref{metric}) and the ansatz $A_\mu=\left(0,A(r),0,0\right)$, the~non-vanishing components of the generalized Riemann curvature tensor are as follows:

\begin{align}
R^0_{101} &= \frac{1}{4} \left(2 A'(r) + (A(r) - \alpha'(r)) (\alpha'(r) - \beta'(r)) - 2 \alpha''(r)\right) \\
R^0_{220} &= \frac{1}{4} e^{-\beta(r)} r (-2 + r A(r)) (A(r) - \alpha'(r)) \\
R^0_{330} &= \frac{1}{4} e^{-\beta(r)} r (-2 + r A(r)) (A(r) - \alpha'(r))\sin^2(\theta) \\
R^1_{001} &= \frac{1}{4} e^{\alpha(r) - \beta(r)} \left(2 A'(r) + (A(r) - \alpha'(r)) (\alpha'(r) - \beta'(r)) - 2 \alpha''(r)\right) \\
R^1_{221} &= \frac{1}{4} e^{-\beta(r)} r \left(-2 (A(r) + r A'(r)) + (-2 + r A(r)) \beta'(r)\right) \\
R^1_{331} &= \frac{1}{4} e^{-\beta(r)} r  \left(-2 (A(r) + r A'(r)) + (-2 + r A(r)) \beta'(r)\right)\sin^2(\theta) \\
R^2_{020} &= R^3_{030} = \frac{1}{4 r} e^{\alpha(r) - \beta(r)} (-2 + r A(r)) (A(r) - \alpha'(r)) \\
R^2_{112} &= R^3_{113} = \frac{1}{4 r}\left(-2 (A(r) + r A'(r)) + (-2 + r A(r))\beta'(r)\right) \\
R^2_{323} &= \frac{1}{4} \left(4 - e^{-\beta(r)} (-2 + r A(r))^2\right) \sin^2(\theta) \\
R^3_{232} &= 1 - \frac{1}{4} e^{-\beta(r)} (-2 + r A(r))^2
\end{align}
\newpage

\subsection[\appendixname~\thesection]{Second Ansatz: $A_\mu=\left(A_0(r),A_1(r),0,0\right)$}

Considering (\ref{metric}) and the ansatz $A_\mu=\left(A_0(r),A_1(r),0,0\right)$, the~non-vanishing components of the generalized Riemann curvature tensor are as follows:

 \begin{subequations}
\begin{align}
R^0_{001} &=R^1_{101}  =R^2_{201} =R^3_{301} =\frac{A_0'(r)}{2} \\
R^0_{101} &=\frac{1}{4} \left(2 A_1'(r) + (A_1(r) - \alpha'(r)) (\alpha'(r) - \beta'(r)) - 2 \alpha''(r)\right) \\
R^0_{220} &=\frac{1}{4} e^{-\beta(r)} r (-2 + r A_1(r)) (A_1(r) - \alpha'(r)) \\
R^0_{221} &=\frac{1}{4} e^{-\alpha(r)} r^2 \left(2 A_0'(r) + A_0(r) (A_1(r) - \alpha'(r))\right) \\
R^0_{330}  &=\frac{1}{4} e^{-\beta(r)} r (-2 + r A_1(r)) (A_1(r) - \alpha'(r)) \sin^2(\theta) \\
R^0_{331}  &=\frac{1}{4} e^{-\alpha(r)} r^2 \left(2 A_0'(r) + A_0(r) (A_1(r) - \alpha'(r))\right) \sin^2(\theta) \\
R^1_{001}  &=\frac{1}{4} e^{\alpha(r) - \beta(r)} \left(2 A_1'(r) + (A_1(r) - \alpha'(r)) (\alpha'(r) - \beta'(r)) - 2 \alpha''(r)\right) \\
R^1_{220}  &=\frac{1}{4} e^{-\beta(r)} r^2 A_0(r) (-A_1(r) + \alpha'(r)) \\
R^1_{221}  &=\frac{1}{4} r \left(-e^{-\alpha(r)} r A_0(r)^2 + e^{-\beta(r)} \left(-2 (A_1(r) + r A_1'(r)) + (-2 + r A_1(r)) \beta'(r)\right)\right) \\
R^1_{313}  &=\frac{1}{4} r  \left(e^{-\alpha(r)} r A_0(r)^2 + e^{- \beta(r)} \left(2 (A_1(r) + r A_1'(r)) + (2 - r A_1(r)) \beta'(r)\right)\right) \sin^2(\theta) \\
R^1_{330}  &= \frac{1}{4} e^{-\beta(r)} r^2 A_0(r) (-A_1(r) + \alpha'(r)) \sin^2(\theta) \\
R^2_{020}  &= R^3_{030}= \frac{1}{4 r}e^{\alpha(r) - \beta(r)} (-2 + r A_1(r)) (A_1(r) - \alpha'(r)) \\
R^2_{021}  &=R^3_{031} = \frac{1}{4} \left(2 A_0'(r) + A_0(r) (A_1(r) - \alpha'(r))\right) \\
R^2_{102}  &=R^3_{103} =\frac{1}{4} A_0(r) (-A_1(r) + \alpha'(r)) \\
R^2_{121}  &= R^3_{131} =\frac{1}{4 r}e^{-\alpha(r) + \beta(r)} r A_0(r)^2 + 2 (r A_1'(r) + \beta'(r)) + A_1(r) (2 - r \beta'(r)) \\
R^3_{232}  &=\frac{1}{4} \left(4 - e^{-\beta(r)} (-2 + r A_1(r))^2 + e^{-\alpha(r)} r^2 A_0(r)^2\right) \\
R^2_{323}  &=\frac{1}{4} \left(4 - e^{-\beta(r)} (-2 + r A_1(r))^2 + e^{-\alpha(r)} r^2 A_0(r)^2\right) \sin^2(\theta) 
\end{align}
\end{subequations}


\end{document}